\documentclass[reprint,amsmath,amssymb, aps,prb]{revtex4-2}

\usepackage{xcolor}
\usepackage{hyperref}
\usepackage{nicefrac}
\hypersetup{pdfborder=0 0 0,colorlinks=true,citecolor=blue,linkcolor=blue}

\usepackage{graphicx}
\usepackage{dcolumn}
\usepackage{bm}

\begin{document}

\title{Simulating the influence of stoichiometry on the spectral emissivity of Mo$_x$Si$_y$ thin films}%

\author{Zahra Golsanamlou}\email{z.golsanamlou@dynamicsolids.net}
 \affiliation{University of Twente, Faculty of Science and Technology and MESA+ Institute 
for Nanotechnology, P.O. Box 217, 7500 AE Enschede, the Netherlands}
\author{Arseniy Baskakov}
 \affiliation{University of Twente, Faculty of Science and Technology and MESA+ Institute 
for Nanotechnology, P.O. Box 217, 7500 AE Enschede, the Netherlands}
\author{Silvester Houweling}

\affiliation{ASML Research, Materials and Chemistry, 5504 DT Veldhoven, the Netherlands}
\author{Giorgio Colombi}
\affiliation{ASML Research, Materials and Chemistry, 5504 DT Veldhoven, the Netherlands}
\author{Robbert van de Kruijs}
 \affiliation{University of Twente, Faculty of Science and Technology and MESA+ Institute 
for Nanotechnology, P.O. Box 217, 7500 AE Enschede, the Netherlands}
\author{Marcelo Ackermann}
 \affiliation{University of Twente, Faculty of Science and Technology and MESA+ Institute 
for Nanotechnology, P.O. Box 217, 7500 AE Enschede, the Netherlands}
\author{Menno Bokdam}\email{m.bokdam@utwente.nl}
 \affiliation{University of Twente, Faculty of Science and Technology and MESA+ Institute 
for Nanotechnology, P.O. Box 217, 7500 AE Enschede, the Netherlands}

\date{\today}

\begin{abstract}
In this work, we simulate the spectral emissivity of various stoichiometric crystal phases of Mo$_x$Si$_y$ compounds using density functional perturbation theory. The dielectric function, including electronic and ionic contributions, is calculated for each phase. We use the bulk properties obtained to simulate the optical absorption spectrum originating from the compound in thin film ($\sim$20~nm) form. We find that most thin films of Mo$_x$Si$_y$ are metallic, however, our results indicate that their emissivity is not simply correlated with the Mo content. For hot metallic films at around 900~K, we predict a maximal emissivity between 5-10~nm thickness. Our results are in good qualitative agreement with experiments, confirming that the emissivity of hexagonal MoSi$_2$ is much lower than in the tetragonal phase. This is related to the small band gap (hexagonal MoSi$_2$) and low density of states at the Fermi level (tetragonal MoSi$_2$). Furthermore, test calculations on defected MoSi$_2$ demonstrate that the infrared emissivity of MoSi$_2$ thin films can be substantially increased by introducing defects.
 
\end{abstract}

\maketitle


\section{Introduction}

Materials that can withstand high temperatures for a long period of time are of increasing scientific and technological importance\cite{Zhang:2018, werner:2001, PAN:2021}. Especially when used as coatings or thin films, their physical and chemical properties on an atomistic scale are important. To prevent overheating, breaking, or melting, a thermal balance has to be established between external heating and cooling of the film. The film can be cooled by thermal conduction to a heat sink, or by passive radiative cooling. The larger open question is how the atomic structure (i.e., crystallinity and defects) of the thin film controls its heat management. Here, we focus on the effect of atomic composition and structure on the dielectric properties of the film. The passive radiative cooling ability of a thin film is expressed by its emissivity; a temperature-dependent quantity with a value between zero and one. An emissivity of one means that the thin film has the same radiative capacity as a black body radiator at temperature $T$, and zero means that at this temperature the film is incapable of emitting any radiation. Film thickness is an important parameter for emissivity. The emissivity of dielectric films decreases with the thickness of the film. For metallic films, the emissivity is constant, but when film-thickness reduces to the order of tens of nanometers, the emissivity suddenly increases as a result of multiple reflections in the thin film\cite{EDALATPOUR:2013}. In this work, we focus on thermal radiation for the case of optically thin films, which means that their thickness is less than the typical penetration depth (also known as skin depth) of the incoming light. We present results from a first-principles based modeling study of thin film emissivity. The presented materials modeling scheme is general, and can aid with the phenomenological interpretation of the experimental observations by simulating the effect of changing atomic structure. We apply the scheme to the case study of Mo$_x$Si$_y$, as our study has been carried out in parallel to recent ellipsometry measurements of Mo$_x$Si$_{1-x}$ thin films by Baskakov~\textit{et al.}\cite{BASKAKOV2025}. Their crystallographic and spectroscopic measurements on sputter-deposited films point towards the importance of crystal phases and electronic structure (band structure and density of states at the Fermi level) combined with the concentration of metal (Mo) in the film.

Mo$_x$Si$_y$ is an interesting, and technologically relevant, class of (polymorphous) thin film materials that can become metallic or semiconducting depending on the stoichiometry. Different crystal phases of Mo$_x$Si$_y$ are reported in the Materials Project data base\cite{merchant:2023, jain:2013}, but only a subset of those have been observed in diffraction experiments (Mo, Mo$_3$Si, Mo$_5$Si$_3$, MoSi$_2$, and Si)\cite{brandes2013,yao:1999}. Crystal structures related to stoichiometries such as MoSi$_3$, MoSi and Mo$_3$Si$_2$ are not thermodynamically stable and phase segregate in stable (poly)-crystalline domains\cite{Czerny:aem24}. Mo$_x$Si$_y$-based materials have elevated melting points, suitable thermal and electrical conductivity, low molecular density and high oxidation resistance, making them suitable for high temperature applications \cite{VASUDEVAN:1992, PAN:2021, DASGUPTA:2008, BAHR:2023, kumar:1993, weis:2016, tao:2022,yang:2023}. Their melting temperature ($\rm T_m$) increases with the atomic fraction of Mo ($\frac{x}{x+y}$), starting at 1687~K for pure Si, increasing to 2293~K for MoSi$_2$, and further to 2453~K for Mo$_5$Si$_3$, upon reaching 2896~K for pure Mo\cite{Gokhale:jope91,Czerny:aem24}. Mo$_x$Si$_y$-based materials are currently being explored for a range of applications, including turbine airfoils, diesel engine glow plugs, combustion chamber components in oxidizing environments, industrial gas burners, molten metal lances, glass processing materials \cite{HAWK:1995, yao:1999} and extreme ultraviolet (EUV) pellicle applications\cite{Choi_2024,Choi_2025}. Tetragonal MoSi$_2$ could specifically be an attractive candidate for display devices, ultraviolet (UV) absorbers, and to prevent solar heating\cite{liton:2022}. For these kinds of applications, thin films of the material are preferred.

Some experimental and theoretical works have already been performed on the emissivity of Mo$_x$Si$_y$-based compounds. Zheng~\textit{et al.}, explored experimentally the potential of pressureless sintered SiC-MoSi$_2$ composites for bulk infrared source applications\cite{Zheng:2019}. They studied the effects of microstructure by increasing the sintering temperature and showed that this results in a decrease of the electrical resistivity and an increase in infrared emissivity.  The very recent experimental investigation of Baskakov~\textit{et al.} on the emissivity of Mo$_x$Si$_{1-x}$ films found that, the emissivity of 20~nm films that were annealed at 600 or 900 $^\circ$C show considerably different emissivities. After annealing at 600$^\circ$C, emissivities between 0.1 and 0.3 are reported for films with Mo content between 17 and 33 at.\%. Annealing at a temperature of 900 $^\circ$C  results in a relative emissivity increase between 10 and 70\%, however not increasing beyond 0.33. They concluded that these differences are related to the crystal phase present in the film, either hexagonal or tetragonal MoSi$_2$.

Previous simulation studies have shown that dielectric and optical properties are related to stoichiometry and crystal structure\cite{vanZetten:prb09, jiao:2011} or quantum confinement\cite{ming:2014}. To date, the emissivity of MoSi$_2$-based compounds has been sparsely investigated with theoretical approaches. Xiang~\textit{et al.} calculated the intrinsic electronic emissivity of tetragonal MoSi$_2$ as 0.17 at 300~K which increases to 0.51 at 1400~K\cite{XIANG:2021}. To increase emissivity, they proposed a change of microstructure by incorporating grains of SiO$_2$. Liton~\textit{et al.} investigated the influence of hydrostatic pressure on the optical and thermal properties of tetragonal MoSi$_2$\cite{liton:2022}. Song~\textit{et al.} investigated the effect of mixing in medium and high-entropy transition metal disilicides on emissivity\cite{SONG2023}.

Apart from MoSi$_2$, the emissivity of thin film Mo$_x$Si$_y$ in other $(\frac{x}{x+y})$ stoichiometric ratios and correspondingly different crystal phases is not well-reported in the literature. In addition, the contribution of the vibrations of the ionic lattice to the dielectric response of the materials, and hence the emissivity, has not been calculated. Therefore, we here explore the emissivity of thin film Mo$_x$Si$_y$ using first-principles and optical methods; including both electronic and ionic contributions, and allowing for multiple reflections within the thin film. A macroscopic optical model is used to simulate the effect of the extrinsic properties, such as film thickness and refractive index of the environment, on the emissivity of the thin film. We characterize the emissivity according to the electronic structure, and we compare the emissivity across material stoichiometry. Finally, we study the effect of structural defects on the emissivity by supercell calculations of MoSi$_2$ test systems.

\section{Computational Methods}

\begin{figure}[b] 
\includegraphics[width =0.98\columnwidth]{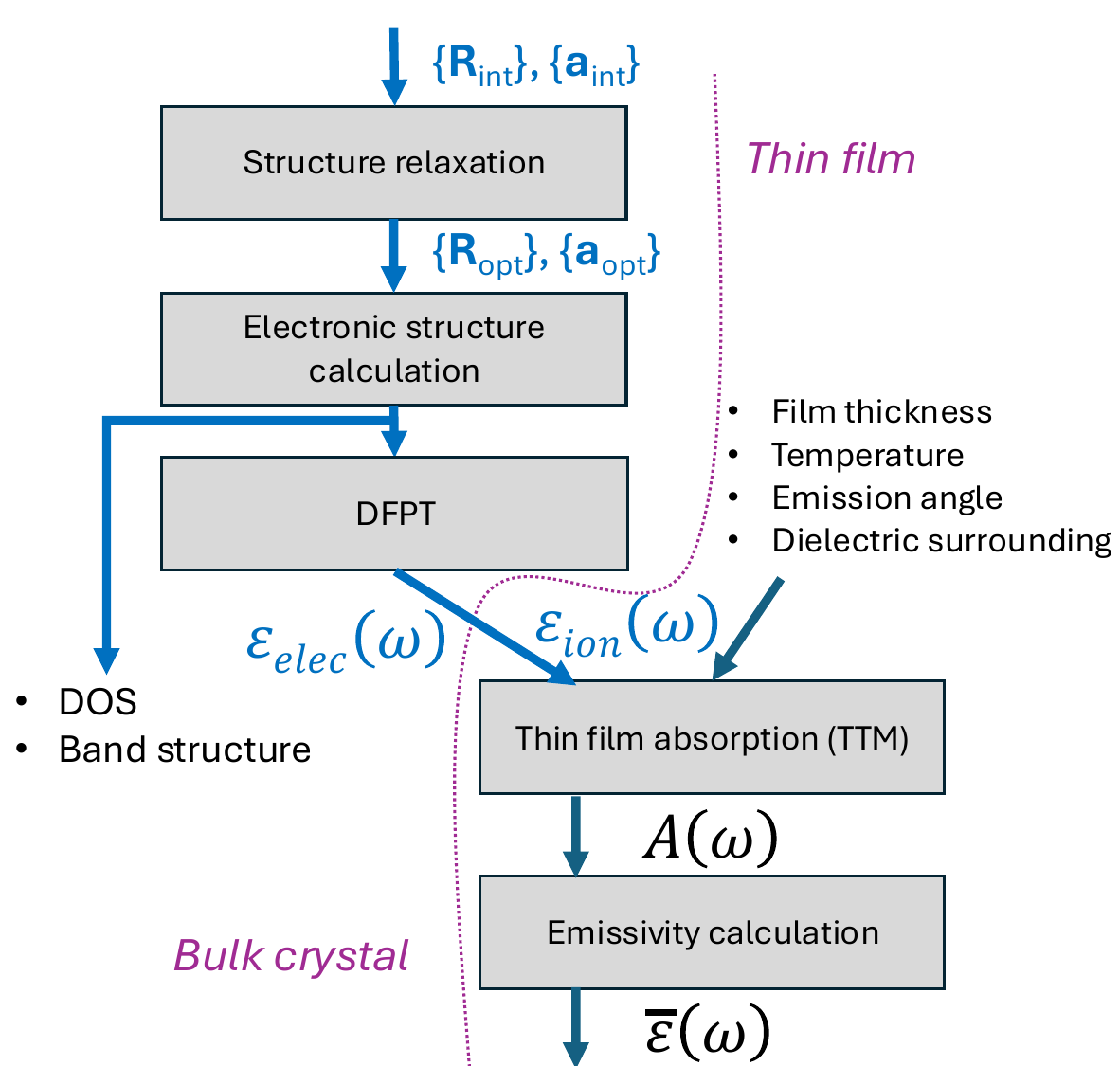}
\caption{Flowchart of the simulation and computation setup to calculate the emissivity of thin films starting from first principles. The input is formed by the atomic coordinates $\{\mathbf{R}_{1\ldots N}\}$ and lattice vectors  $\{\mathbf{a}_{1,2,3}\}$ of the structures in Figure~\ref{fig:struc}. The purple dotted line divides the chart, the left hand is on the level of bulk crystal and the right hand side on the level of the thin film.}
\label{fig:flow}
\end{figure}

A flowchart of our simulation approach is drawn in Figure~\ref{fig:flow}. We start with a calculation of the electronic structure at the level of the bulk crystal. The resulting dielectric functions are passed on to calculate the absorption and emissivity of the material at the level of the thin film. In the following two subsections, we elaborate, in order of appearance from top to bottom, on the various methods in the flowchart.

\subsection{Bulk crystal}
The first-principles calculations presented in this work are based on Density Functional Theory (DFT). We consider several different crystal phases of Mo$_x$Si$_y$ and optimize the related geometries until the total energy change between successive iterations is below $10^{-6}$~eV. The DFT calculations are performed using the Projector Augmented Wave (PAW) method\cite{paw-1} as implemented in the Vienna Ab-initio Simulation Package (VASP)\cite{Kresse:prb93,Kresse:prb96,paw-2} using a plane-wave basis with a cutoff of 520~eV. The generalized gradient approximation (GGA) of PBEsol\cite{perdew:prl2008} is used as the exchange-correlation functional. The Brillouin zone integration is performed on $\Gamma$-centered Monkhorst-Pack k-point meshes, and a Gaussian smearing width of $\sigma=0.2$~eV broadens the one-electron levels. The same $\sigma$ value is used for all materials and corresponds to a relatively high temperature broadening. For density of states (DOS) calculations, the tetrahedron method is used on the final relaxed structures. A $18\times18\times6$ k-point mesh is used to sample the Brillouin zone of tetragonal MoSi$_2$. For the other Mo$_x$Si$_y$ structures k-point meshes of similar density are used, taking into account their respective lengths of the lattice vectors and thereby creating the mesh with the closest matching k-point sampling density to that of MoSi$_2$.\newline

\noindent The dielectric function is modeled as the addition of independent terms on a common frequency grid
\begin{equation}
 \begin{aligned}
    \varepsilon(\omega)&=\;\;\;\;\;\;\;\;\;\;\;\;\varepsilon_{\rm elec.}(\omega)&+\;\;\varepsilon_{\rm ion}(\omega) \\
    &=\overbrace{\varepsilon_{\rm inter}(\omega)+\varepsilon_{\rm intra}(\omega)}&+ \;\;\varepsilon_{\rm ion}(\omega). 
    \end{aligned}\label{eq:epsilon}
\end{equation}
We create an isotropic solution by averaging the diagonal of the dielectric tensor $\varepsilon_{\alpha\beta}(\omega)$, where $\alpha,\beta$ are the Cartesian directions. For the electronic part, we distinguish between inter- and intra-band contributions. The inter contribution is calculated using Density Functional Perturbation Theory (DFPT) within the density-density picture\cite{Gonze:prb97,Wu:prb05,Gajdos:prb06}. For intrinsic semiconductors and insulators, all valence bands are fully occupied and (small) excitations inside the band cannot be made, resulting in an intra-band contribution of zero. For metals and doped or small band gap semiconductors at finite temperatures, the intra-band transitions can be described by a free-electron plasma (Drude) model, where the dielectric function is parameterized by the plasma frequency ($\omega_{\rm p}$) and a damping factor ($\gamma$)\cite{LeeKeun-HoChang:prb94,Harl:prb07}. The intra-band dielectric function composed of real and imaginary parts is then written as\cite{vanZetten:prb09}
\begin{equation}
    \varepsilon_{\rm intra}(\omega)=\varepsilon_{\infty}-\frac{\omega^2_p}{\omega^2+\gamma^2}+\frac{\gamma\omega_p^2}{\omega^3+\omega\gamma^2}i,
    \label{eq:drude}
\end{equation}
where the constant $\varepsilon_{\infty}$ describes the screening by electronic inter-band transitions. In a pure Drude metal $\varepsilon_{\infty}$ is the vacuum permittivity of 1, but in the Mo$_x$Si$_y$ materials there are inter-band transitions making $\varepsilon_{\infty}=\lim_{\omega\to 0}\varepsilon_{\rm inter}(\omega)$. When adding the intra-contribution to the common grid (Eq.~\ref{eq:epsilon}) we set $\varepsilon_{\infty}$ in Eq.~\ref{eq:drude} to zero in order to prevent double counting the vacuum permittivity. The plasma frequencies are computed for the eight crystalline structures in this study based on the full DFT calculated electronic dispersion near the Fermi level. In Supplementary Material (SM) \cite{supp} Section 1, the calculation method for $\omega_{\rm p}$ and its dependence on sufficiently dense k-point grids and the applied smearing width, is discussed. The damping  factor is treated as a parameter. Throughout, a value of $\gamma=0.15$~eV is set, which is between the values obtained in experiment\cite{BASKAKOV2025}, and is a typical value for metals. For the ionic part, the contributions are computed using the $\mathbf{q}=0$ harmonic phonon modes and the Born effective charge tensors, as implemented in VASP\cite{Bokdam:sr16}. 

\subsection{Thin film}
The optical penetration depth ($d_{\rm pen}$) is a measure of how far light can penetrate inside a material. We use it to assess whether multiple reflections inside the thin film are to be expected, i.e., when the film thickness $d\le d_{\rm pen}$. The penetration depth is calculated as
\begin{equation}
    d_{\rm pen}=\frac{\lambda(\omega)}{4\pi k(\omega)},
\end{equation}
where the wavelength of the incoming light $\lambda(\omega)$ is expressed in nm and $k(\omega)$ is the extinction coefficient. Both the refractive coefficient $n(\omega)$ and the extinction coefficient $k(\omega)$ are calculated from the real/complex parts of the bulk dielectric function $\varepsilon(\omega)=\varepsilon'(\omega)+i\varepsilon''(\omega)$ as follows
\begin{equation}
\begin{split}
n(\omega)&=\sqrt{\frac{|\varepsilon(\omega)|+\varepsilon'(\omega)}{2}}\\
k(\omega)&=\sqrt{\frac{|\varepsilon(\omega)|-\varepsilon'(\omega)}{2}}\;.
\end{split}
\end{equation}
The approximation of the dielectric properties of a thin film by its bulk counterpart starts to fail when the film thickness is of the order of the Fermi wavelength. Quantum confinement effects in metals typically start below $\sim 10$~nm. A direct DFPT calculation of the dielectric response of Mo$_x$Si$_y$ thin film slabs is possible, but is not done here. Such slab calculations have been done for Al(111)\cite{ming:2014} and Pb(111)\cite{Zubizarreta:prb17}. Compared to Al or Pb ($\omega_{\rm p}\approx$ 10~eV), the onset of quantum size effect for Mo$_x$Si$_y$ ($\omega_{\rm p}\approx$ 2~eV) is expected at smaller film thicknesses. 

Since $d_{\rm pen}$ is in most cases much larger than the thickness of the film, multiple reflections inside the film are possible and will alter the absorption of the material in the thin film compared to the bulk form. Therefore, we apply the Transfer Matrix Method (TMM) as implemented in the AbsorptionTMM code \cite{byrnes:arxive2020, TMMabs}. The $n(\omega)$ and $k(\omega)$ enter the TTM as parameters. With the TTM we describe the effect of multiple reflections and the varying incident angle of incoming $p$-polarized radiation on the absorption of the thin film of thickness $d$ surrounded on the top and bottom by air ($n=1$). A schematic view of wave propagation in the film is represented in Fig.~S2 in SM.  The calculated absorption is a function, $A=1-T-R$ of the transmission $(T)$ and the reflection $(R)$ of the thin film.

\begin{figure*}[t] 
\includegraphics[scale=0.4]{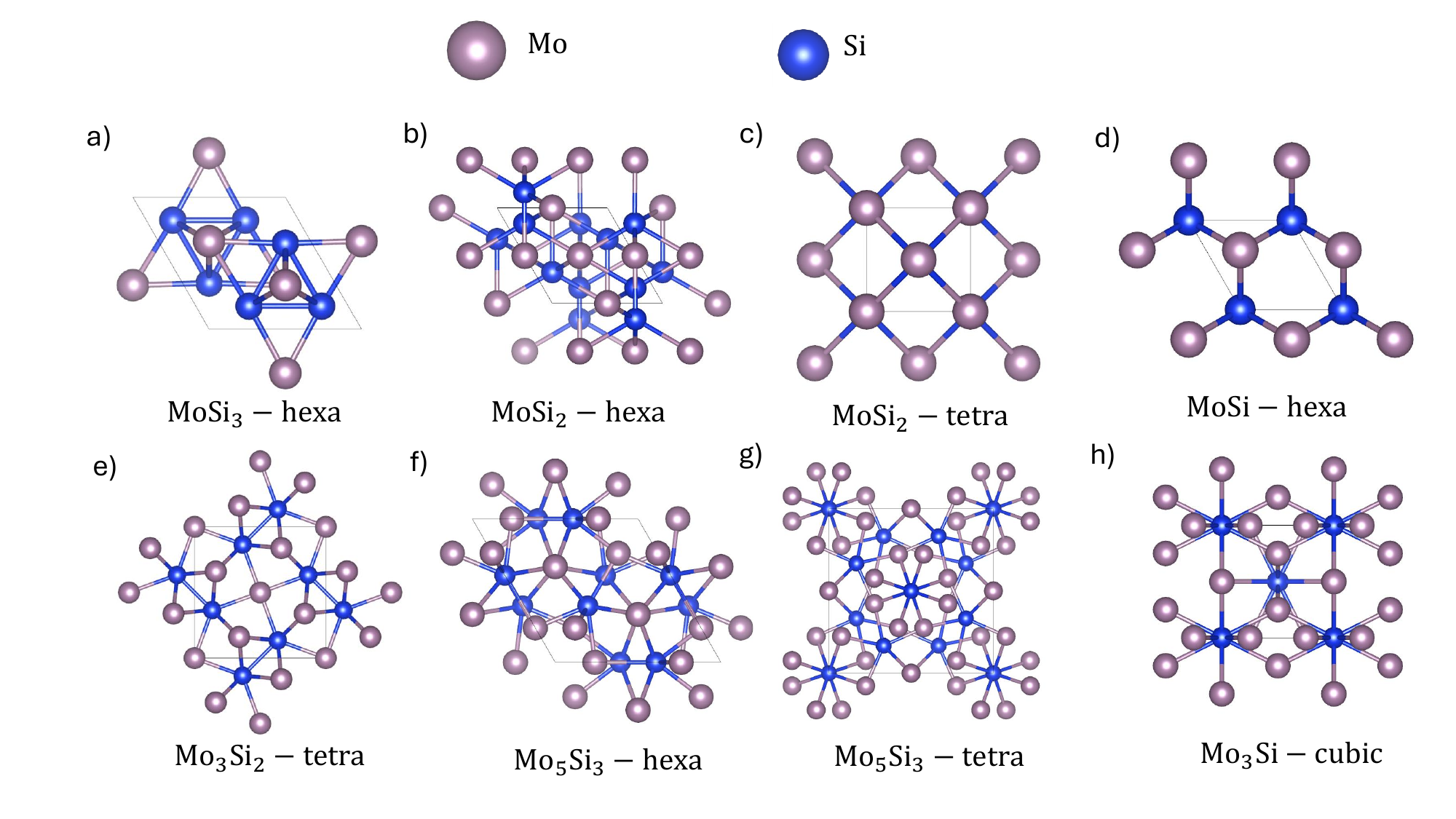}
\caption{Overview of the unit cells considered for the different phases of Mo$_x$Si$_y$:  a) hexagonal MoSi$_3$, b) hexagonal MoSi$_2$, c) tetragonal MoSi$_2$, d) hexagonal MoSi , e) tetragonal Mo$_3$Si$_2$, f) hexagonal Mo$_5$Si$_3$, g) tetragonal Mo$_5$Si$_3$, and h) cubic Mo$_3$Si.} 
\label{fig:struc}
\end{figure*}

\section{Results}
\begin{figure*}[t] 
\hspace{.6cm}\includegraphics[width =0.94\columnwidth]{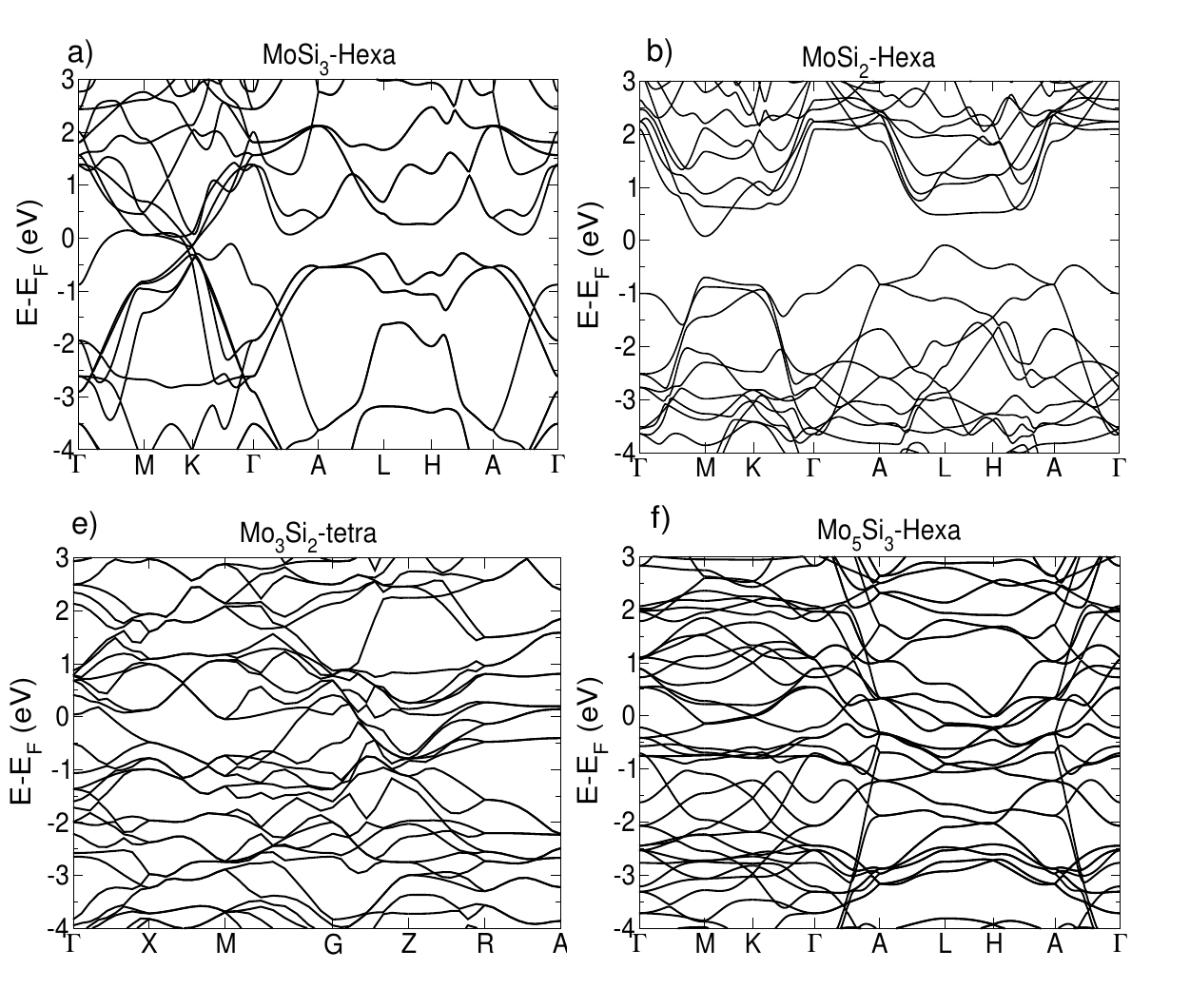}
\includegraphics[width =0.95\columnwidth]{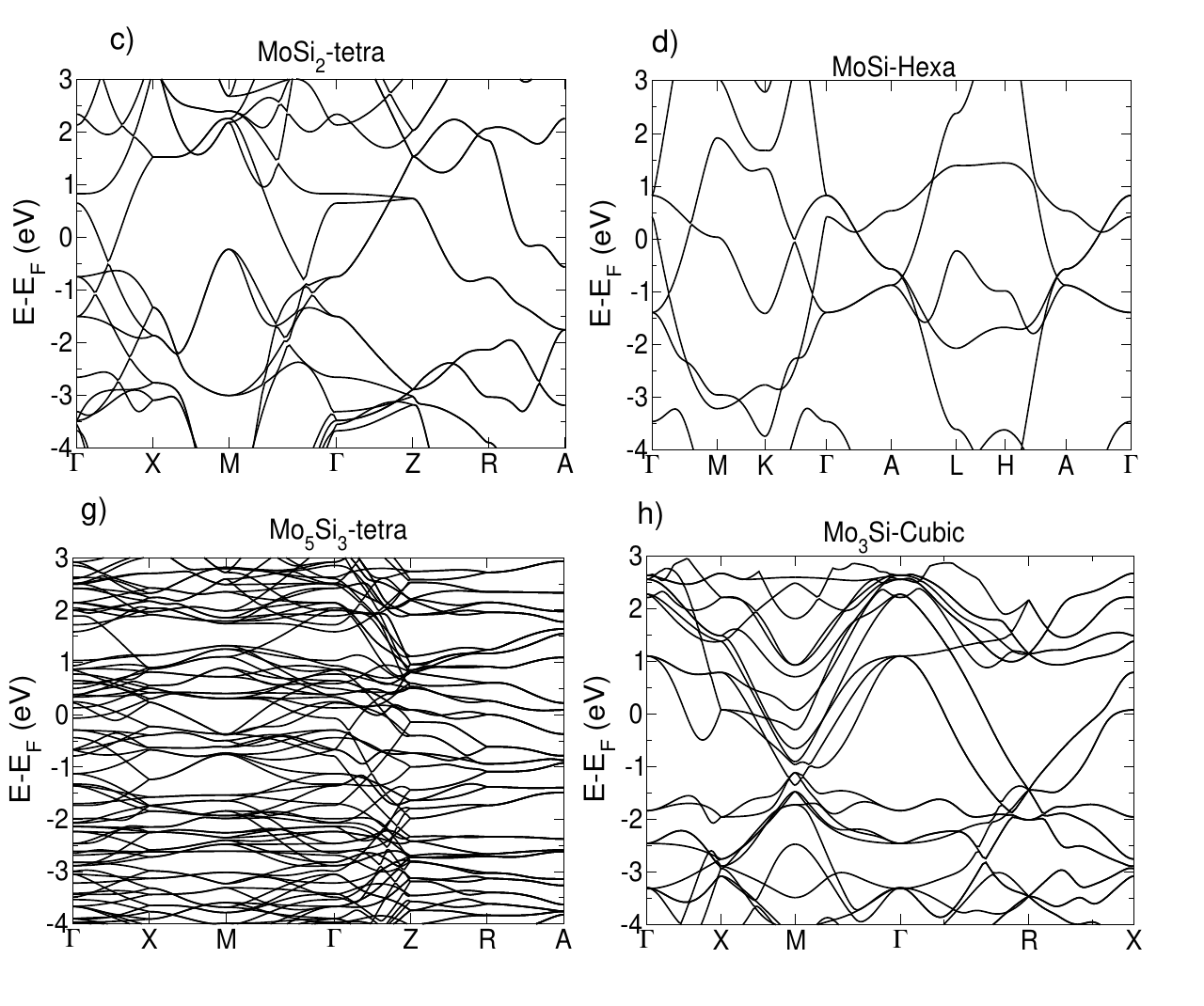}
\includegraphics[width =0.915\columnwidth]{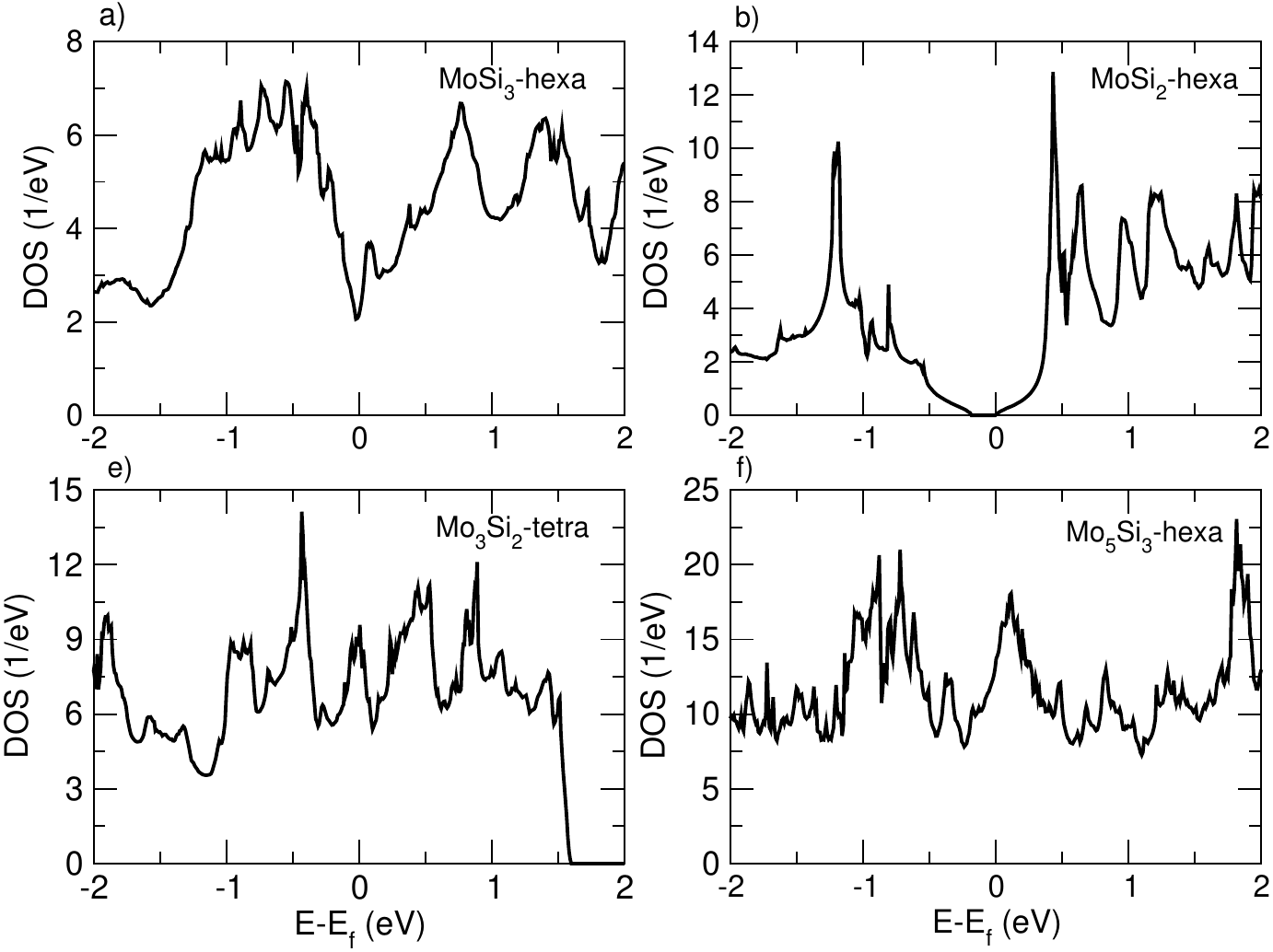}\hspace{4mm}
\includegraphics[width =0.89\columnwidth]{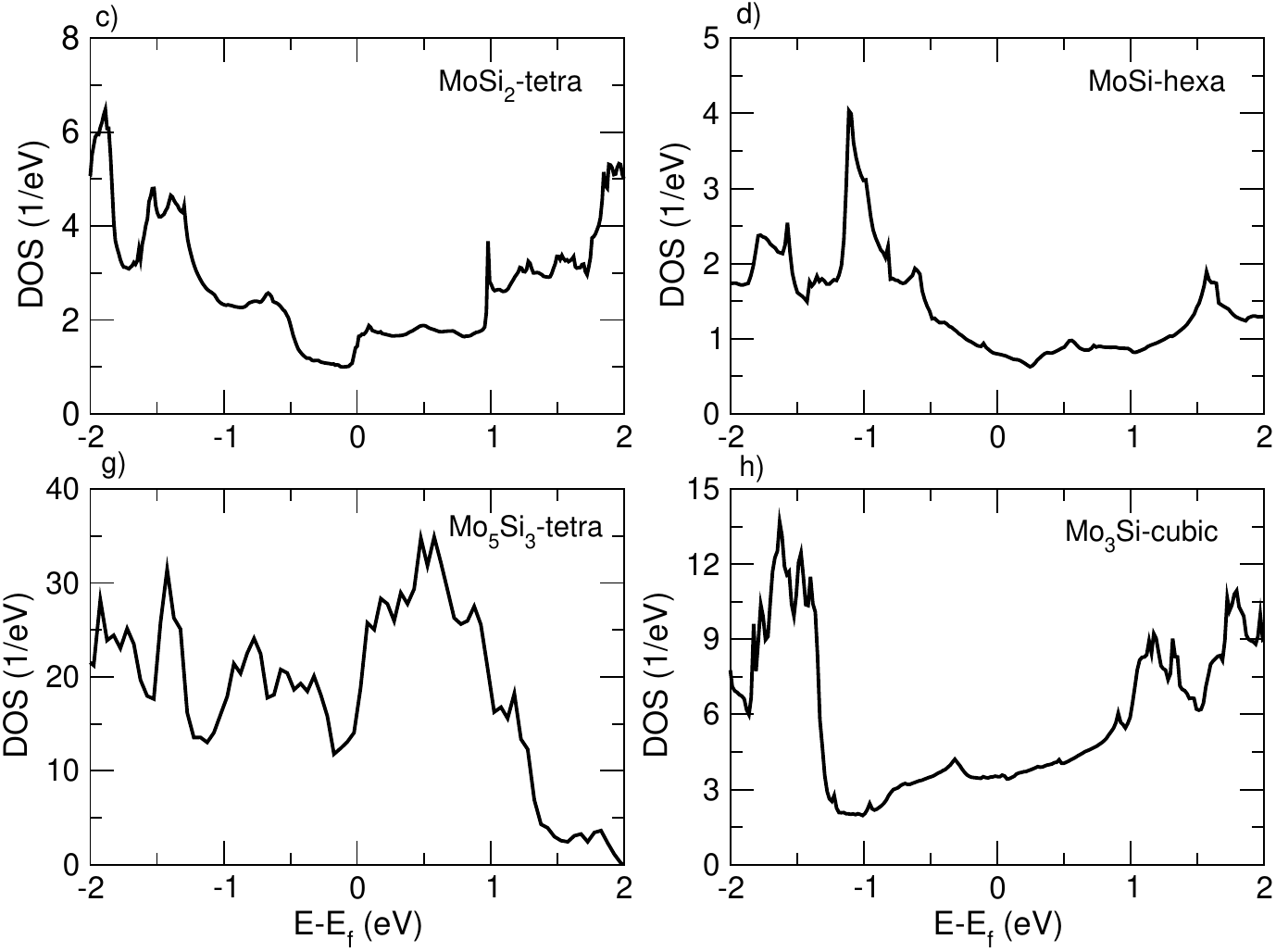}
\includegraphics[width =0.95\columnwidth]{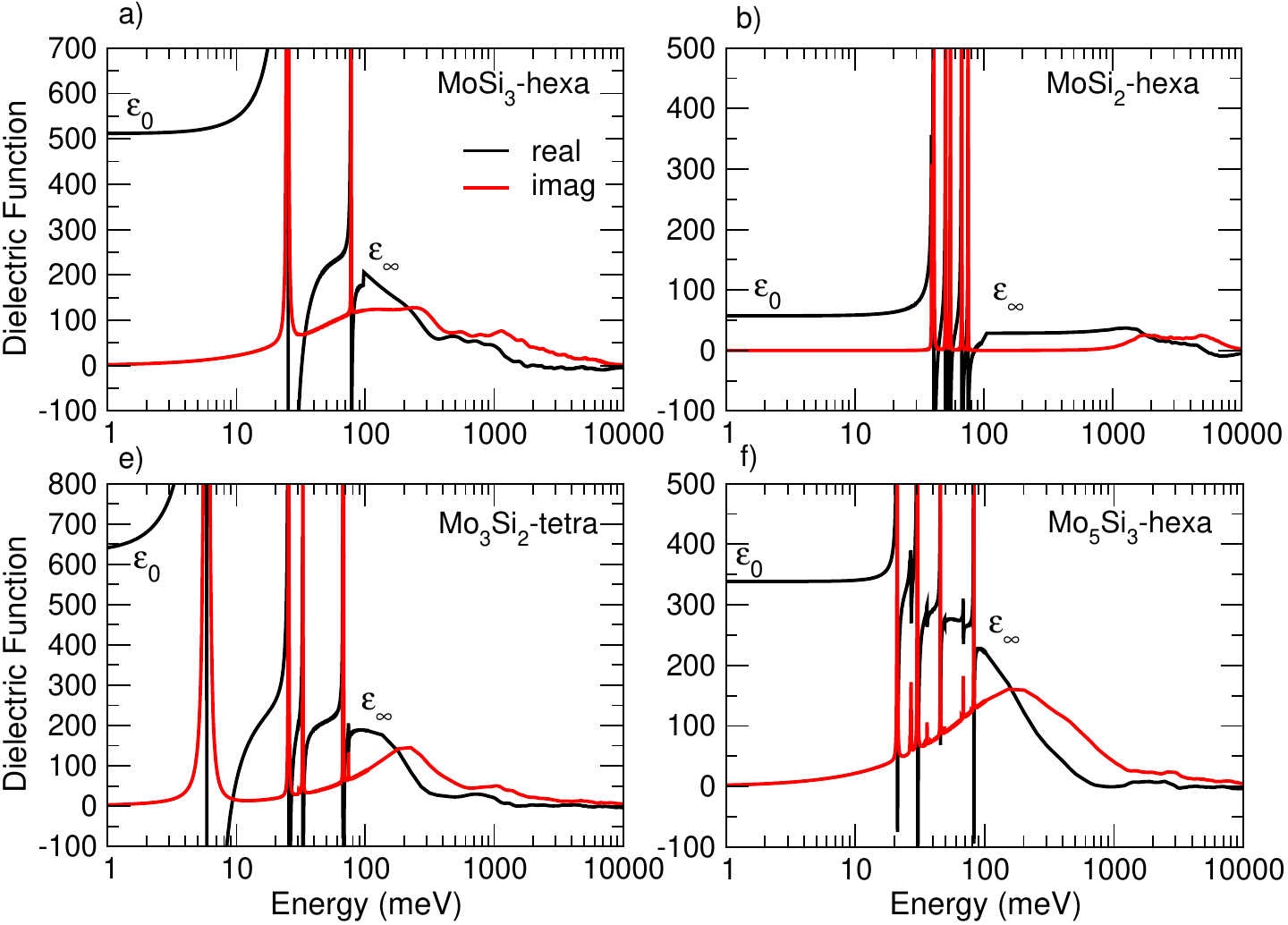}
\includegraphics[width =0.95\columnwidth]{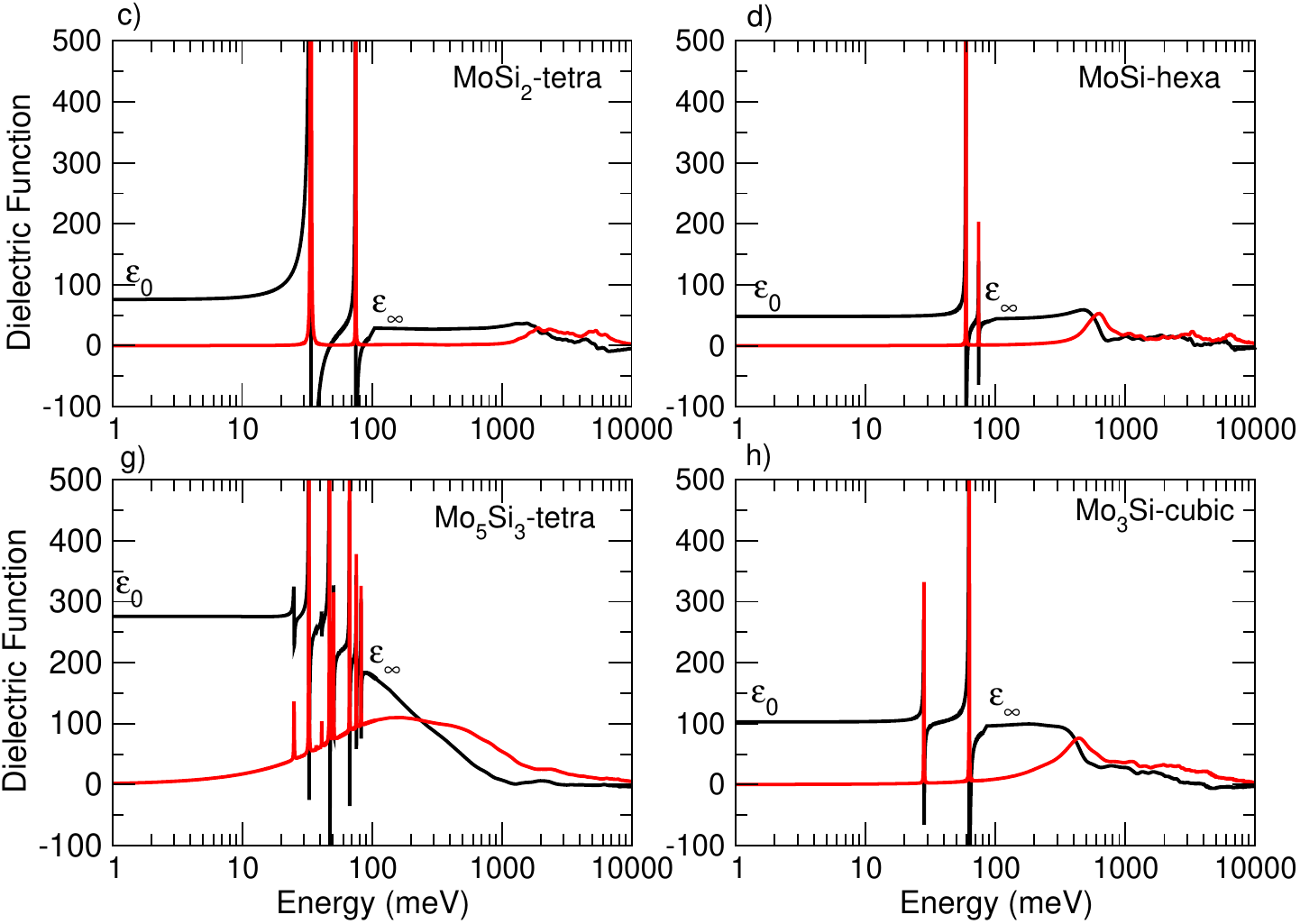}
\caption{From top to bottom: Band structure, density of states, and dielectric function of a) hexagonal MoSi$_3$, b) hexagonal MoSi$_2$, c) tetragonal MoSi$_2$, d) hexagonal MoSi, e) tetragonal Mo$_3$Si$_2$, f) hexagonal Mo$_5$Si$_3$, g) tetragonal Mo$_5$Si$_3$, and h) cubic Mo$_3$Si. The dielectric function is based on electronic inter-band and ionic contributions only. The black/red line denotes the real/imaginary part in $\varepsilon(\omega)=\varepsilon'(\omega)+i \varepsilon''(\omega)$, respectively.} 
\label{fig:bandsdosdiel}
\end{figure*}

Figure~\ref{fig:struc} shows the atomic structures of the different crystal phases of Mo$_x$Si$_y$ that are considered in this work. Five experimentally known stable crystal structures are included: MoSi$_2$ (tetragonal and hexagonal), Mo$_5$Si$_3$(tetragonal and hexagonal), and  Mo$_3$Si (cubic). For both MoSi$_2$ and Mo$_5$Si$_3$ the tetragonal system is the stable low temperature structure, and MoSi$_2$ is reported to transition to a metastable hexagonal structure around 2173~K\cite{Gokhale:jope91}. Additionally, we include crystal structures obtained by first-principles simulations\cite{osti_1315941,Woods:matter23} that are most likely not thermodynamically stable: MoSi$_3$ (hexagonal), MoSi (hexagonal), and Mo$_3$Si$_2$ (tetragonal), but enable us to study dielectric properties over the stoichiometric range of $0.25 \le\frac{x}{x+y}\le 0.75$.

Before simulating the emissivity of Mo$_x$Si$_y$ in thin-film form, we first examine the electronic structure of the crystals. In Figure~\ref{fig:bandsdosdiel}~(top row), the band structures along high-symmetry lines\cite{Setyawan:ComMatSci2010} of the eight crystal structures are shown. We see that all, except for one, of the crystals are metals with a finite density of states (DOS) at the Fermi level ($\rm E_F$). Only MoSi$_2$ in the hexagonal phase is a small band-gap semiconductor. In Figure~\ref{fig:bandsdosdiel}~(middle row), the DOS($\rm E$) of the Mo$_x$Si$_y$ crystals are shown over an energy range $\pm2$~eV around $\rm E_F$. A high DOS($\rm E_F$) is an indication that the material is potentially a good metal with high conductivity. In the Drude model, a high DOS($\rm E_F$) can be seen as a high valence electron density $\rho_v$. This model only describes intra-band electronic screening and links $\rho_v$ to the plasma frequency by $\omega_{\rm p}=\sqrt{\frac{\rho_ve^2}{\epsilon_0 m}}$, with electron charge $e$ and effective mass $m$, and vacuum permittivity $\epsilon_0$. As the imaginary part of the Drude dielectric function Eq.~(\ref{eq:drude}) scales with $\omega_{p}^2$, the bulk absorption coefficient 
($\alpha=\frac{2\omega{}}{c}k$) is a function of $\rho_v$ through the extinction coefficient $k(\rho_v,\omega)$. We have tabulated the DOS($\rm E_F$) per volume of the unit cell in Table~\ref{tab:1}. The DOS($\rm E_F$) is not the same throughout the seven metallic systems, therefore, it is expected that the conductivity of the films will vary with Mo$_x$Si$_y$ stoichiometry, provided that these variations are not exactly compensated by differences in effective mass and relaxation time.
Being that Mo carries six valence electrons, and Si only four, one might expect that increasing Mo/Si ratio in a Mo$_x$Si$_y$ film leads to a higher DOS nearby the $\rm E_F$. However, for the calculated parameters tabulated in Table~\ref{tab:1}, we do not find a statistically significant correlation (p-value of 0.07) between the stoichiometric ratio and DOS($\rm E_F$), indicating that the electronic properties of Mo$_x$Si$_y$ films cannot be trivially inferred by stoichiometry only. The influence of the crystal structure is further shown in the following, where we calculate dielectric function, absorption spectrum, the plasma frequency (determined by the curvature of the full band structure around $\rm E_{F}$), and emissivity.

\begin{table}[!h]
        \centering    
   \caption{Calculated materials parameters of Mo$_x$Si$_y$ crystals:  DOS($\rm E_F$)/$\Omega$ ($\rm 1/(eV\r{A}^3)$), static dielectric constant $\varepsilon_0$, high frequency dielectric constant $\varepsilon_\infty$, and Drude plasma frequency $\omega_{\rm p}$.}
        \begin{tabular}{ | c | c | c | c | c | c |}
                \hline
      $\frac{x}{x+y}$ &  Crystal & DOS($\rm E_F$)/$\Omega $ & $\varepsilon_\infty$ &$\omega_p (eV)$& $\varepsilon_0$  \\
                \hline
        0.25 &$\rm MoSi_3$ (hexa) & 0.019& 512 & 206&  4.17  \\
        0.33 &$\rm MoSi_2$ (hexa) & & 57 & 290&  0.50  \\
         0.33 &$\rm MoSi_2$ (tetra) & 0.029& 76 & 29& 2.83  \\
        0.5 &$\rm MoSi$ (hexa)& 0.027 & 48 & 44 &  3.81 \\
         0.60 &$\rm Mo_3Si_2$ (tetra) & 0.06& 642 & 177& 4.04  \\
        0.63 & $\rm Mo_5Si_3$ (hexa)& 0.055  & 339 & 169& 3.11  \\
        0.63 &$\rm Mo_5Si_3$ (tetra) & 0.06& 262 & 155&  3.19  \\
         0.75 &$\rm Mo_3Si$ (cubic)& 0.03 & 103 & 96  & 5.57\\     
                \hline 
        \end{tabular} 
   \label{tab:1}
\end{table}

\subsection*{Bulk dielectric function}

To calculate optical properties of the materials, we need to calculate the frequency-dependent response of the material to an electric field. Here, we calculate the dielectric function Eq.~(\ref{eq:epsilon}) of the electronic system and in the ionic lattice, separately. The DFPT calculated dielectric functions, including electronic inter-band and ionic contributions, for all eight Mo$_x$Si$_y$ crystals are presented in Figure~\ref{fig:bandsdosdiel}~(bottom). The real and imaginary parts of the dielectric function are colored black and red, respectively. In the high-frequency range only electronic contributions are present, whereas the ionic contributions (here represented by sharp features) dominate below $\sim100$~meV. The phonon frequencies of the lattice determine the frequency range of the ionic contributions. Only 'dipole-active' phonon modes have finite oscillator strength and contribute to the dielectric function. Fig.~\ref{fig:bandsdosdiel}~(bottom) shows that the ionic contributions have a sizable effect at low response frequencies, resulting in a static dielectric constant $\varepsilon_0$ that is substantially higher than the ‘ion-clamped’ high frequency dielectric constant $\varepsilon_\infty$, see Table~\ref{tab:1}. 

The effect of intra-band screening by conduction electrons is approximated with a Drude model. As input for this model, we have used a fixed damping factor $\gamma=0.15$~eV and calculated the plasma frequency parameter ($\omega_{\rm p}$) for the eight crystals and tabulated them in the last column of Table~\ref{tab:1}. Compared to $\omega_{\rm p}$ values obtained from fits of the spectra measured in Ref.~\cite{BASKAKOV2025} the here obtained numbers are $\sim$3-4 times higher. We identify two main possible explanations for the large difference. Firstly, the small thickness of the film has a decreasing effect on the effective plasma frequency and an increasing effect on the damping factor as compared to the bulk\cite{Shah2022}. Secondly, the lower level of crystallinity of the Mo$_x$Si$_y$ film grown in the experiment, compared to the perfect crystals in the calculations. Unfortunately, a direct DFT calculation of the plasma frequency of an amorphous or polymorphous structure currently lies (because of the required system size) outside the available computational capabilities. Therefore, we will hereafter present results with and without the intra-band contribution, and analyze the emissivity of a pure Drude thin film over a range of plasma frequencies.

\subsection*{Optical penetration depth \& absorption}

\begin{figure}[!t]
\includegraphics[width =\columnwidth]{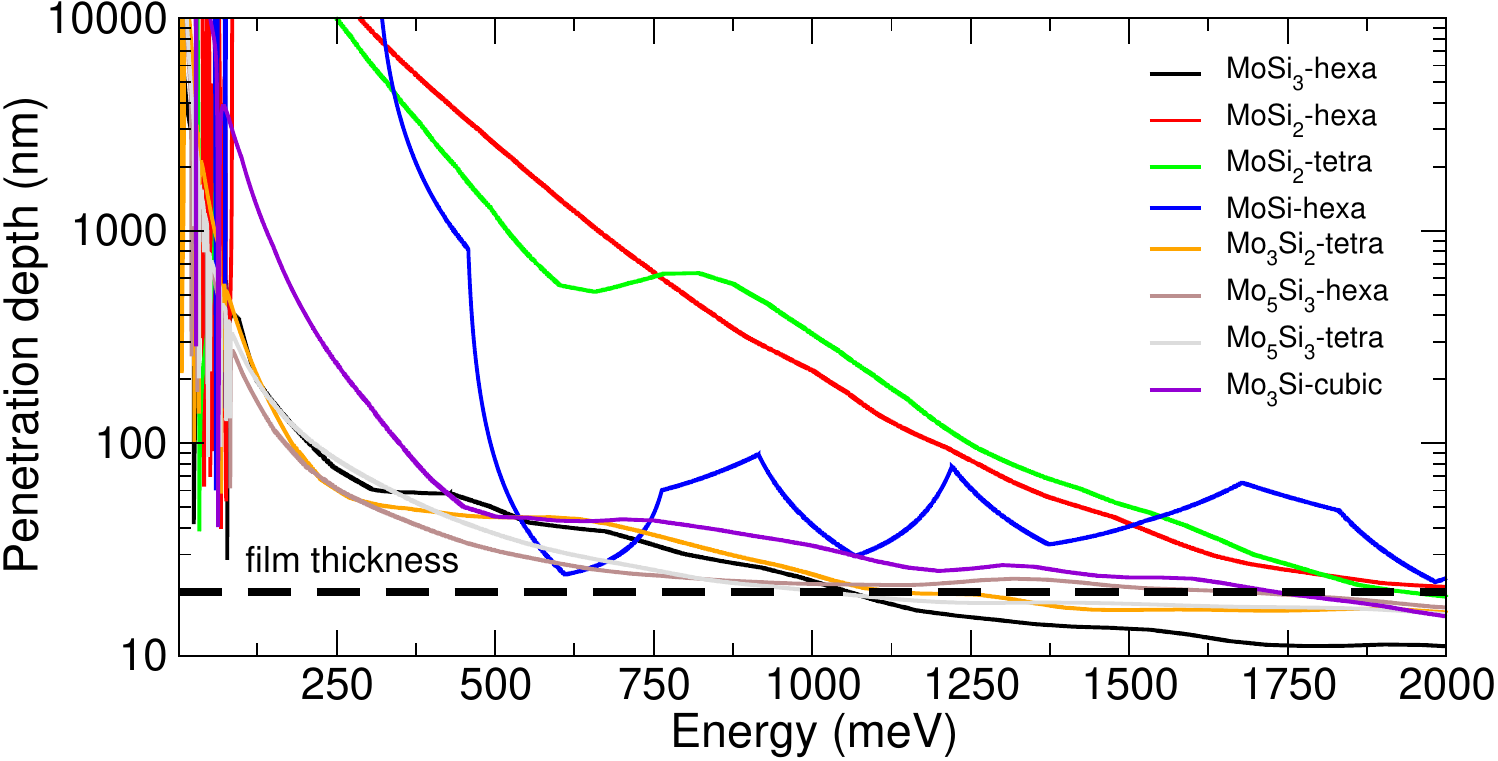}
\includegraphics[width =\columnwidth]{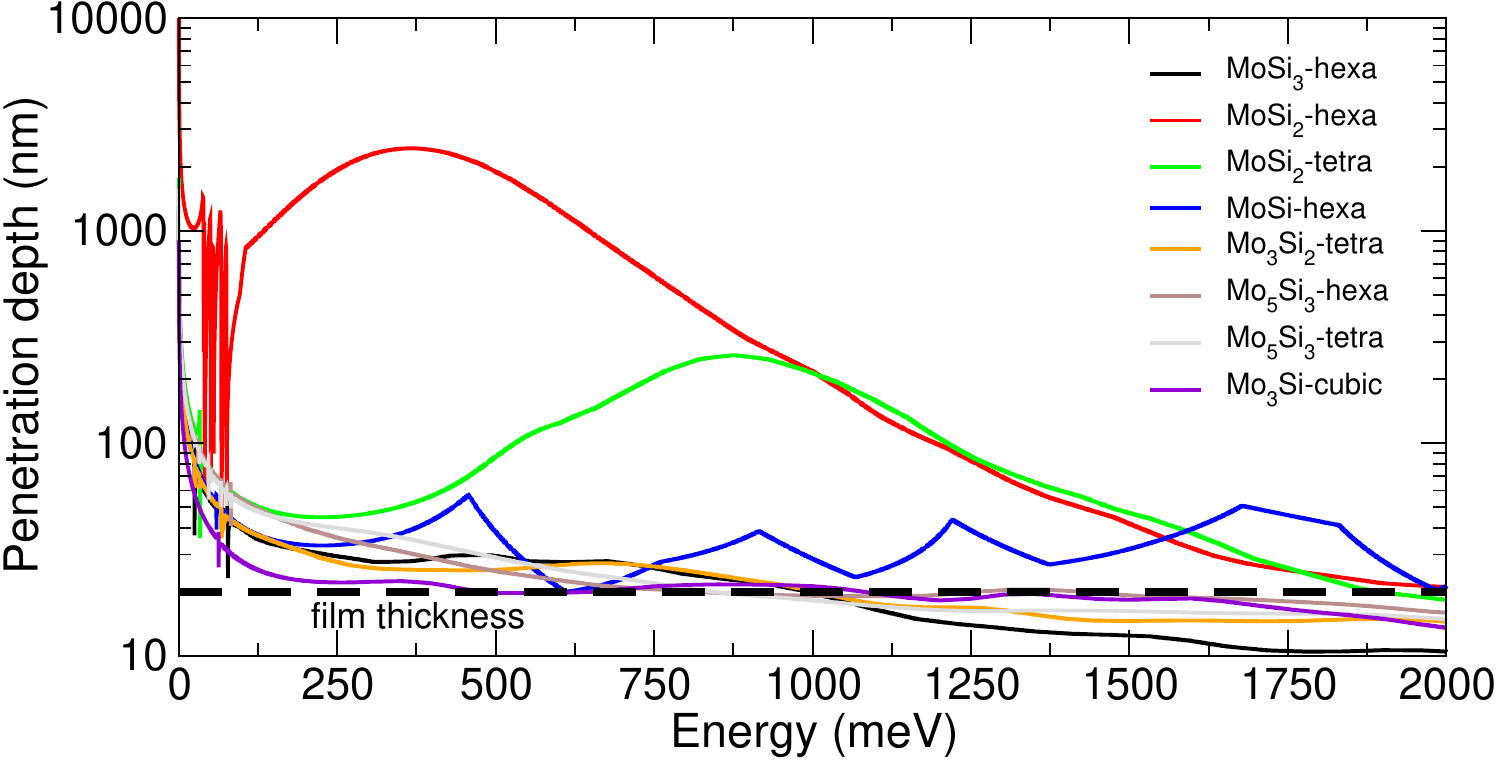}
\caption{The frequency-dependent penetration depth of the eight Mo$_x$Si$_y$ materials based on (\textit{top}) only the inter+ion dielectric function of Fig.~\ref{fig:bandsdosdiel}, and the (\textit{bottom}) full intra+inter+ion dielectric function of Eq.~\ref{eq:epsilon}) The typical thin film thickness is indicated by the horizontal dashed line.} 
\label{fig:pend}
\end{figure}

We have calculated the frequency-dependent penetration depth ($d_{\rm pen}$) of the eight materials based on only the inter and ion contributions to the bulk dielectric function. It is the length scale after which the intensity of monochromatic light that has entered the material has decayed by a factor of $\frac{1}{e}$. The resulting functions $d_{\rm pen}(\omega)$ for the eight materials are shown in Figure~\ref{fig:pend}. We find that the film thickness ($d$) (typically $\sim$20~nm) is either $d<d_{\rm pen}$ or, for some materials, even $d\ll d_{\rm pen}$ across most of the infrared (IR) spectrum ($\sim0.01-1.7$~eV).  This makes the absorption and emission of these materials in thin films different from those of the bulk material. A fraction of the light transmitted through the first air$|$Mo$_x$Si$_y$ interface will reflect (internally) at the Mo$_x$Si$_y|$air interface, and another fraction will be transmitted. This process repeats itself many times inside the film. Including additional electronic intra-band contributions causes the penetration depth to decrease, as shown in Figure~\ref{fig:pend}~(bottom). Optical absorption is enhanced especially in the low-frequency range. Still, $d<d_{\rm pen}$ in the part of the spectrum relevant for emissivity, as we will see when calculating the absorption spectrum. Therefore, the film incorporating all three contributions to the dielectric function should also be regarded as an (optical) thin film. As a consequence,  a description of reflection and absorption by a single Fresnel equation at the air$|$Mo$_x$Si$_y$ interface will not capture the effect of internal reflections. Therefore, the absorption of the thin film is computed using the TTM method, which accounts for multiple reflections. The computed bulk dielectric function is used as input for the TTM method and results in an absorption $A(\omega,\theta,d)$ that is a function of frequency and incident angle of the light, and film thickness, respectively. For the calculation it is assumed that the film is freestanding, surrounded at the top and bottom by vacuum ($\varepsilon=1$), and that the incoming light is p-polarized and hits the surface under normal incidence $\theta=90^\circ$. See SM Fig.~S2 for a schematic overview.

\begin{figure}[!t]
\includegraphics[width=\columnwidth]{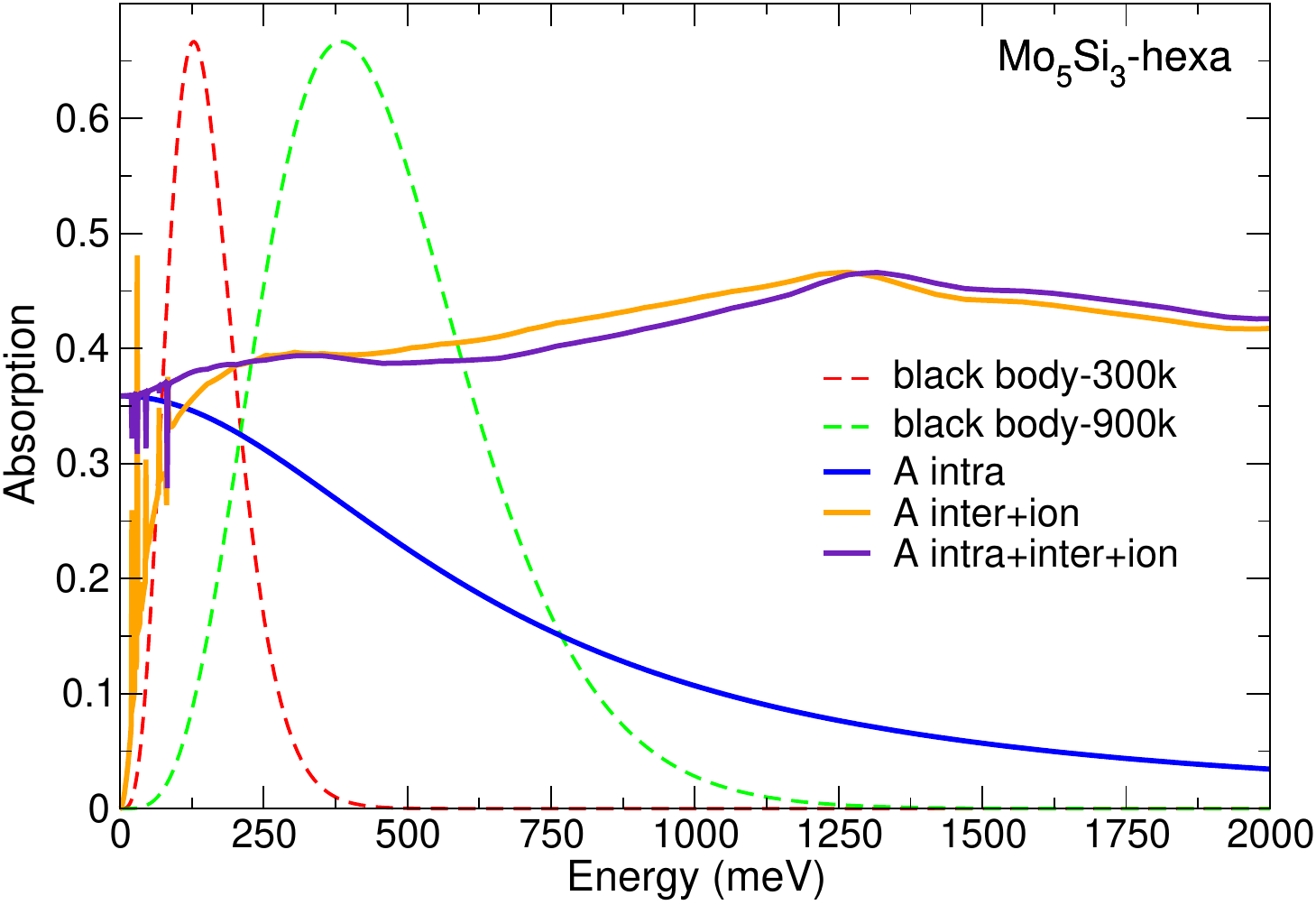}
\includegraphics[width=\columnwidth]{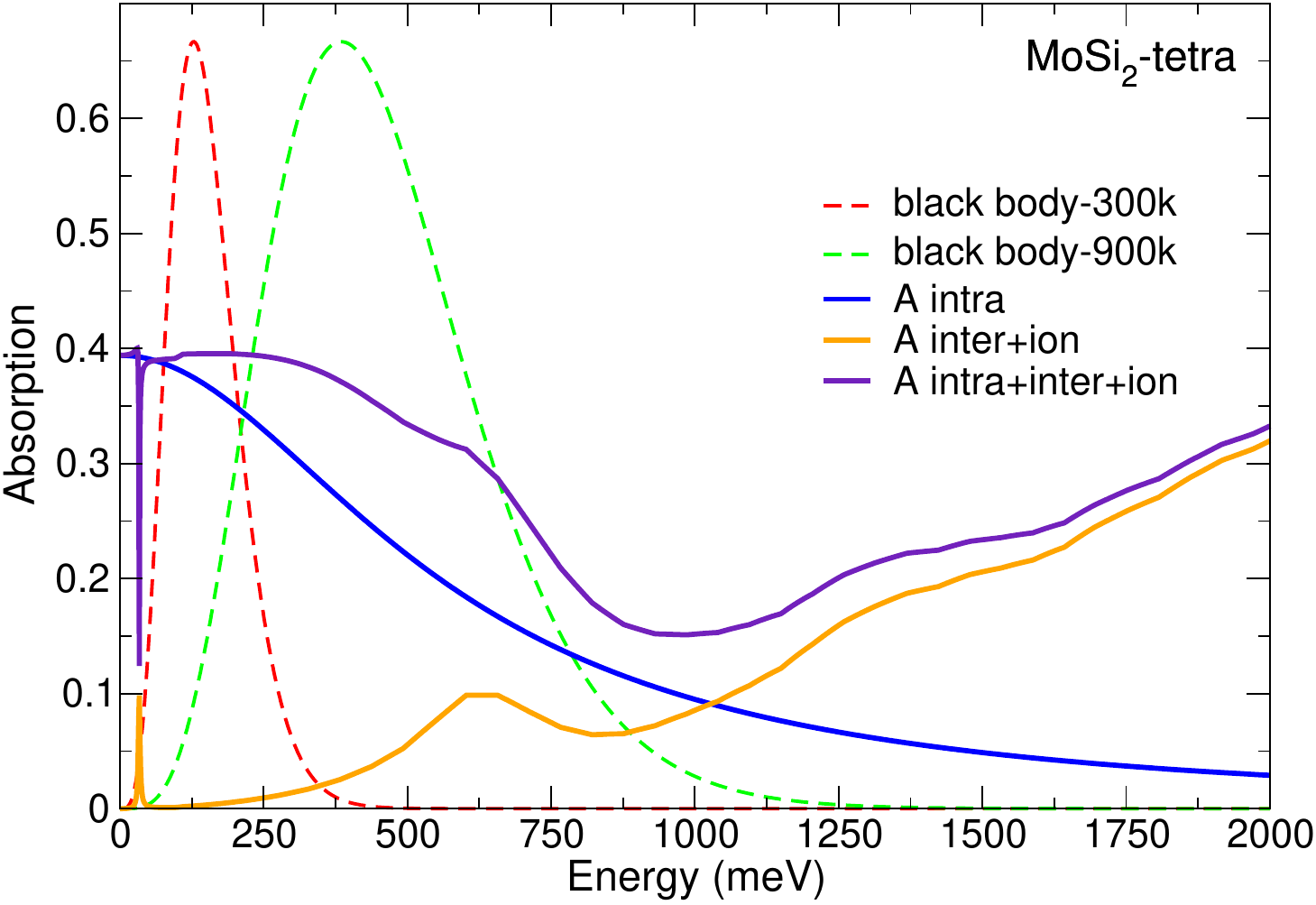}
\caption{Absorption spectra for a 20~nm (\textit{top}) hexagonal Mo$_5$Si$_3$ and (\textit{bottom}) tetragonal MoSi$_2$ thin film including different dielectric contributions, and black body radiation spectra at temperatures of 300~K and 900~K. (Black body curves are normalized by peak height.)} 
\label{fig7:abs}
\end{figure}

We will now examine the absorption spectra $A(\omega)$ of a 20~nm thin film under normal incidence angle of two representative materials; one with low penetration depth in the low IR domain, viz. hexagonal Mo$_5$Si$_3$ and one with high penetration depth, viz. tetragonal MoSi$_2$. The calculated spectra are shown in Figure~\ref{fig7:abs} . The three different absorption spectra show the effect of including the different contributions. The for emissivity relevant part of the spectrum, should have overlap with the spectral radiance of a black body. The black body spectrum is described by Planck's law and shown in the two figures for temperatures of 300 and 900~K. By raising the temperature, the black body spectrum covers a different part of the absorption spectra, resulting in a different emissivity. We see that the absorption by intra-band electrons is important at low(er) temperature and becomes less important when the film gets hotter. The contribution of electronic inter-band transitions then starts to dominate. We also see that the overlap with the ionic contributions, i.e. the sharp spikes typically below $\sim$50~meV, shows already at 300~K little overlap with the black body spectrum, indicating a minor contribution to the emissivity at this temperature. Comparing the inter+ion spectra to the full intra+inter+ion absorption spectra, we see that for hexagonal Mo$_5$Si$_3$ the differences are minor whereas they are large for tetragonal MoSi$_2$. This means that uncertainties related to the $\omega_{\rm p},\Gamma$-parameters in the Drude description will have a larger effect on the calculated emissivity for systems such as hexagonal and tetragonal MoSi$_2$, but will only play a minor role for the other Mo$_x$Si$_y$ films. The absorption spectra for the other six Mo$_x$Si$_y$ stoichiometries are added in the SM (Fig.~S4).

\subsection*{Thin film emissivity}

 \begin{figure}[t] 
\includegraphics[width=\columnwidth]{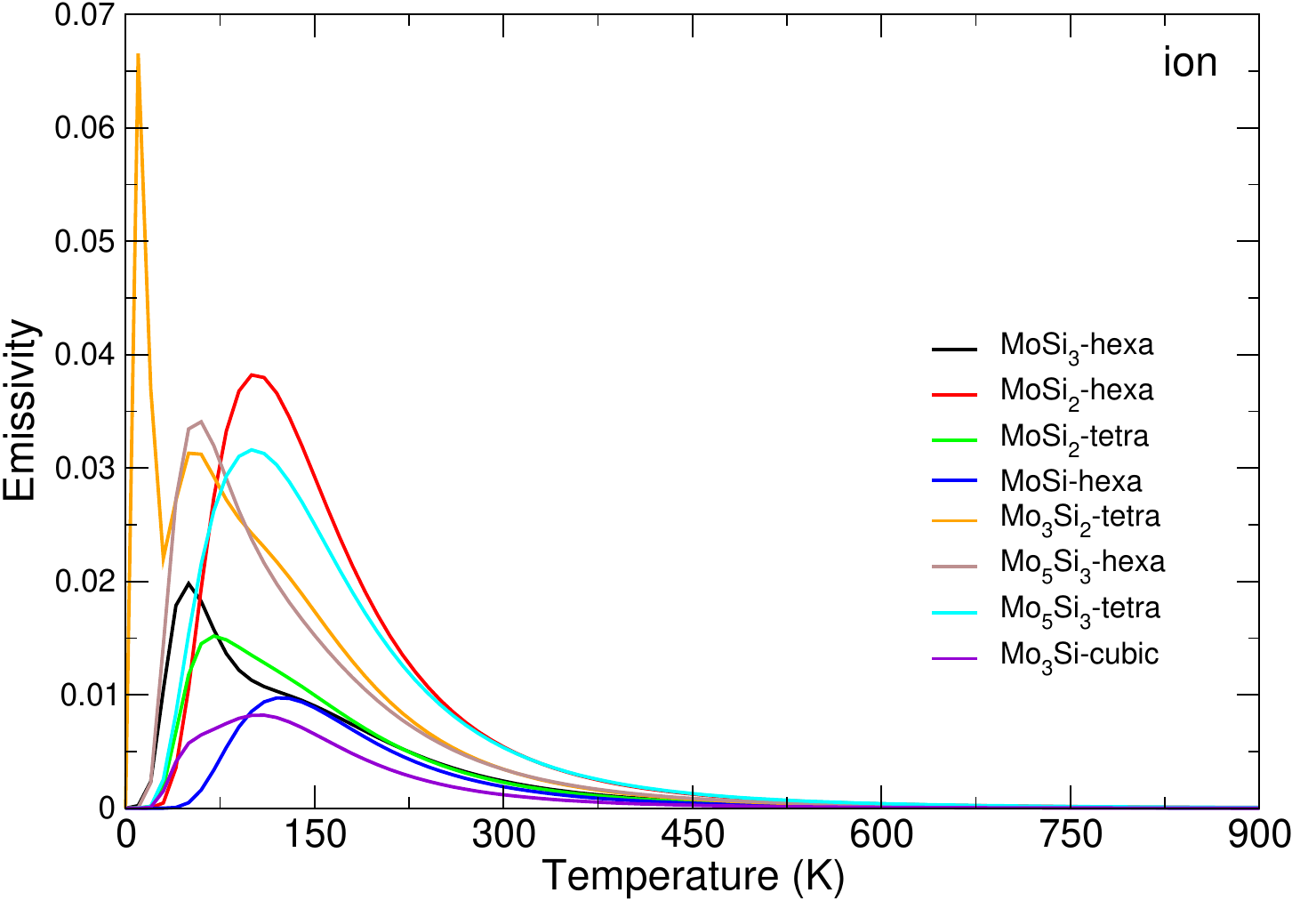}
\includegraphics[width=.987\columnwidth]{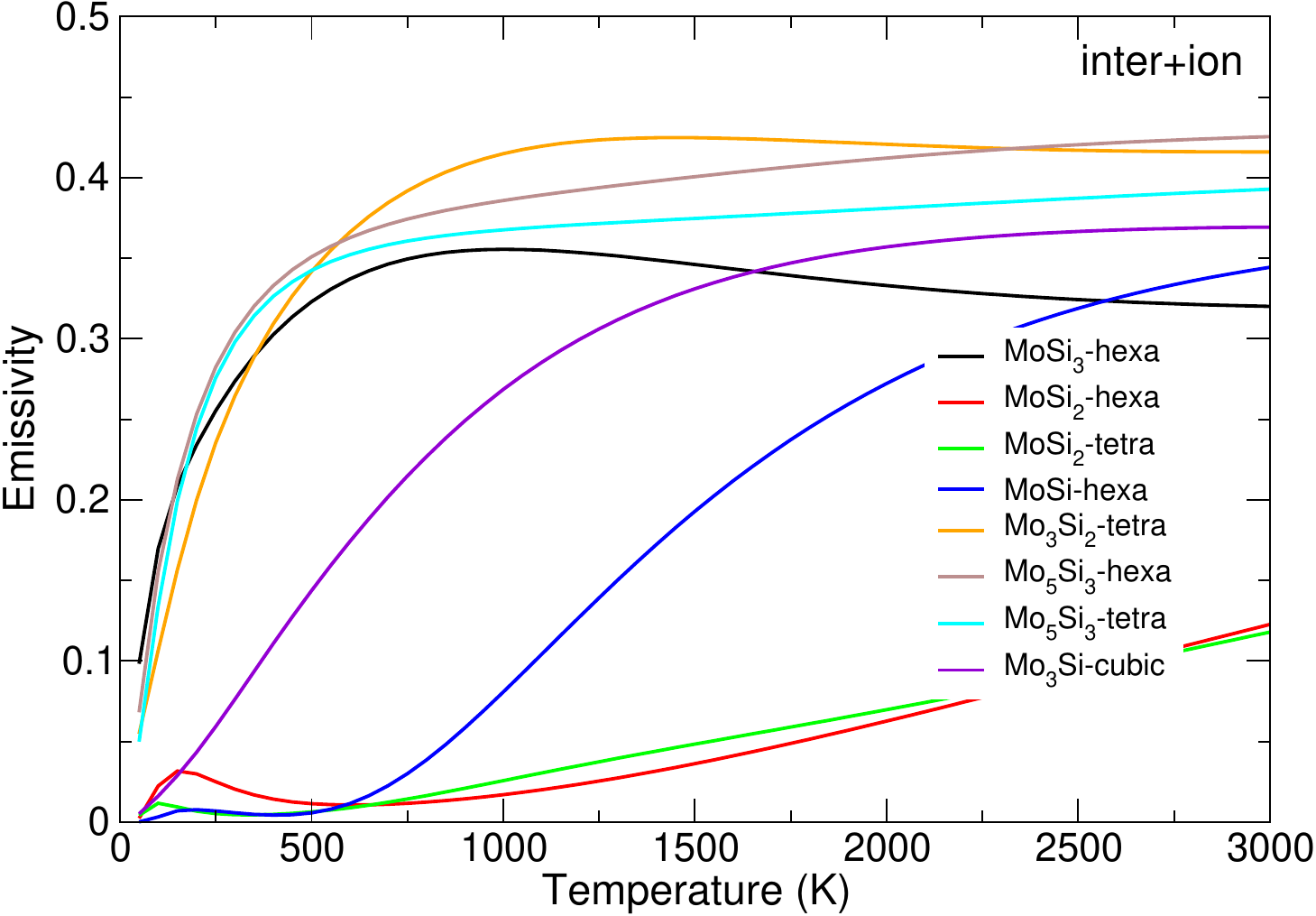}
\includegraphics[width=.987\columnwidth]{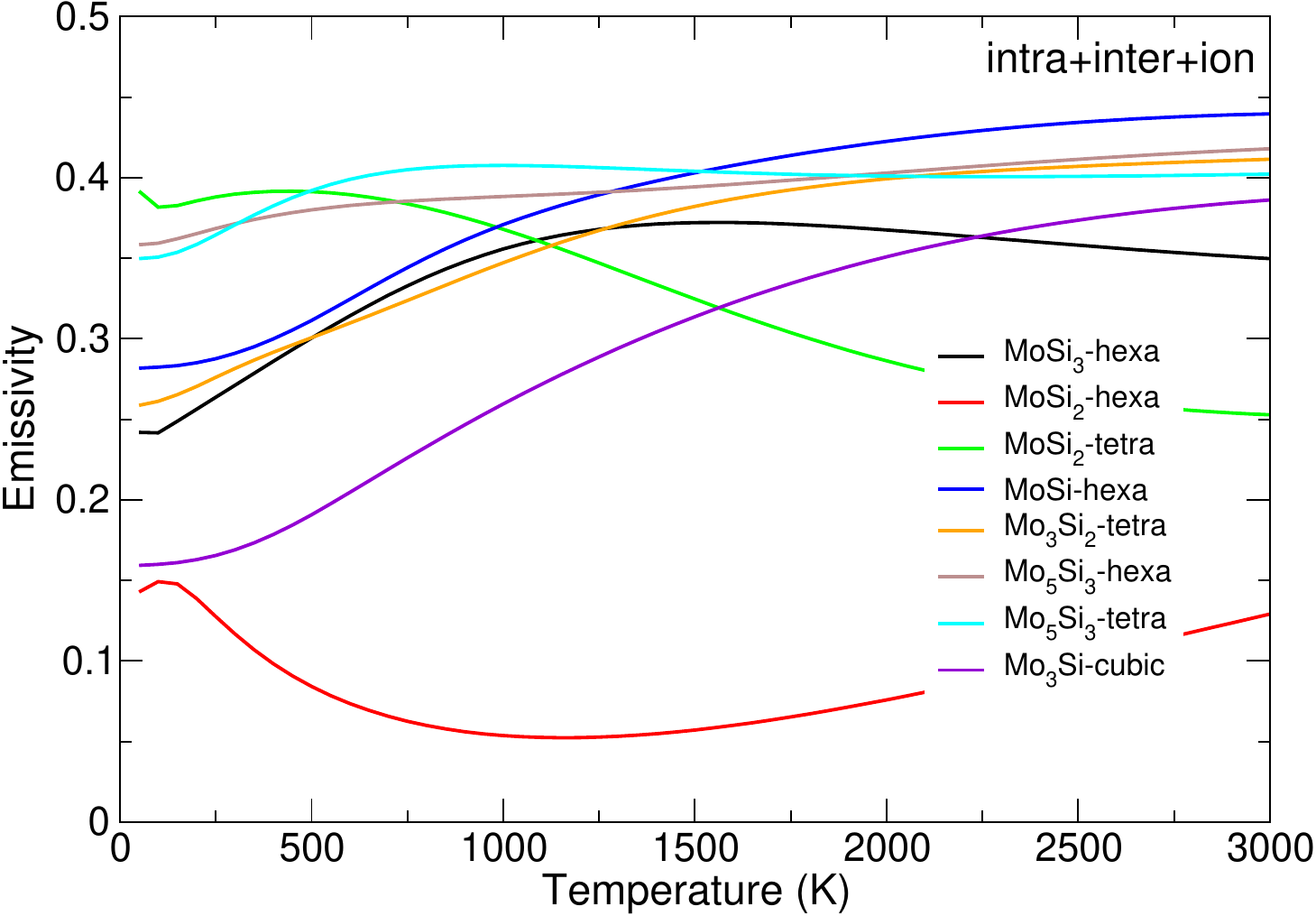}
\caption{Temperature dependence of the emissivity of 20~nm thin film Mo$_x$Si$_y$ based on the (\textit{top}) ionic, (\textit{middle}) inter-band + ionic, and (\textit{bottom}) all three contributions.} 
\label{fig10:emis}
\end{figure}

The thermal emittance of the Mo$_x$Si$_y$ crystals in thin film form is quantified by the emissivity:
\begin{equation}
    \bar{\varepsilon}(T,\theta)=\frac{\int A(\omega,\theta,d)B(\omega,T)d\omega }{\int B(\omega,T)d\omega},
\label{eq:emiss}
\end{equation}
which is a number between zero and one\cite{Siegel:book01}. Kirchhoff's law is applied in Eq.~(\ref{eq:emiss}), meaning that to maintain the thermal equilibrium of the film, its emissivity is equal to its absorptivity. $B(\omega,T)$ is the spectral radiance of a black body at temperature $T$. In this setup both the top and bottom surfaces emit the same power. The temperature dependence of the emissivity for all eight crystalline films is shown in Figure~\ref{fig10:emis}. The 0 to 3000~K temperature range is evaluated, irrespective of whether the crystal is thermodynamically stable or molten at these temperatures. Note that the figure scales on the y-axes differ, and that the three separate contributions to the dielectric function Eq.~(\ref{eq:epsilon}) are added incrementally (ion, ion+inter, ion+inter+intra) from top to bottom in the three plots.

\begin{figure}[!t] 
\includegraphics[width=\columnwidth]{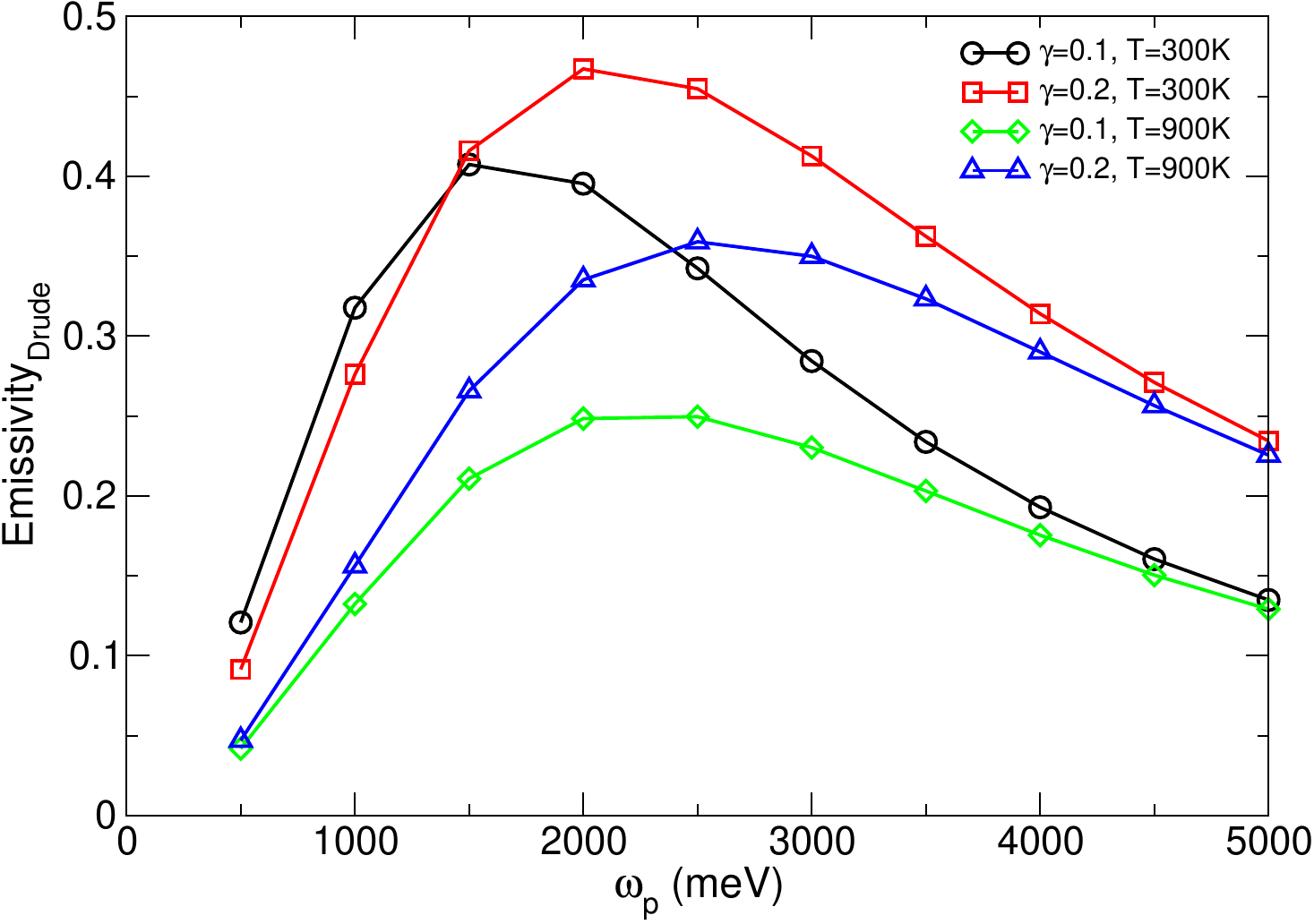}
\caption{Emissivity of a 20~nm Drude thin film in which the intra-band electronic excitations are parameterized by the plasma frequency ($\omega_{\rm p}$) and a damping factor $(\gamma)$.} 
\label{fig13:drude-test}
\end{figure}

As discussed in the previous section, the contribution of ionic screening to the emissivity is expected to be low. Therefore, we have calculated the emissivity solely from the ionic contributions and present the results in Figure~\ref{fig10:emis}~(top). The contribution of the IR emission from the ionic lattice is indeed low; depending on the material, the lattice contributes between 3 to 10 percent of the total emissivity. Above room temperature (RT), the ionic contribution to the emissivity becomes even negligible. MoSi$_2$-hexa forms the exception here, because of its small electronic contribution to emissivity as we will see next. In Figure~\ref{fig10:emis}~(middle), the inter-band electronic contributions are added next to the ionic contributions, resulting in an increase of the emissivity by an order of magnitude. For all materials the emissivity increases with temperature to a high value, however, without the intra-band contribution both phases of MoSi$_2$ show much lower emissivity. In the last Figure~\ref{fig10:emis}~(bottom), the intra-band electronic contributions are added as well. This substantially increases the emissivity at low temperatures for all materials, whereas the differences are minor at high temperatures. Surprisingly, the film with the highest emissivity at RT, MoSi$_2$-tetra, is the least emissive in Fig.~\ref{fig10:emis}~(middle), i.e. without including the intra-band contribution. To understand this, we calculate the emissivity of a pure Drude thin film. Figure~\ref{fig13:drude-test} shows four Drude emissivity curves; for two typical values of the damping factor (slightly above and below $\gamma=0.15$) and at 300 and 900 ~K. We see that intra-band contribution to the thin film emissivity can be substantial and depends sensitively on the plasma frequency. The calculated value of $\omega_{\rm p}$ for MoSi$_2$-tetra (see Table~\ref{tab:1}) lies close to the peak emissivity. 
We note that this emissivity is already close to the theoretical maximum of 0.5, determined by the absorption of a thin film in air occurring when $n=k$\cite{Hadley:47}. The increase around RT for MoSi$_2$-hexa is the result of a small finite $\omega_{\rm p}$ value, which is a more valid approximation in the high temperature regime. Given that we compute a bandgap of 0.2~eV, $\omega_{\rm p}=0$, ie. no intra-band contribution would be a better approximation at RT. Summarizing, we can conclude that the emissivity in Mo$_x$Si$_y$ films predominantly arises from electronic contributions to the dielectric function, and that they do not show a simple (linear) relation with the stoichiometric ratio.

\begin{figure}[!t] 
\includegraphics[width=\columnwidth]{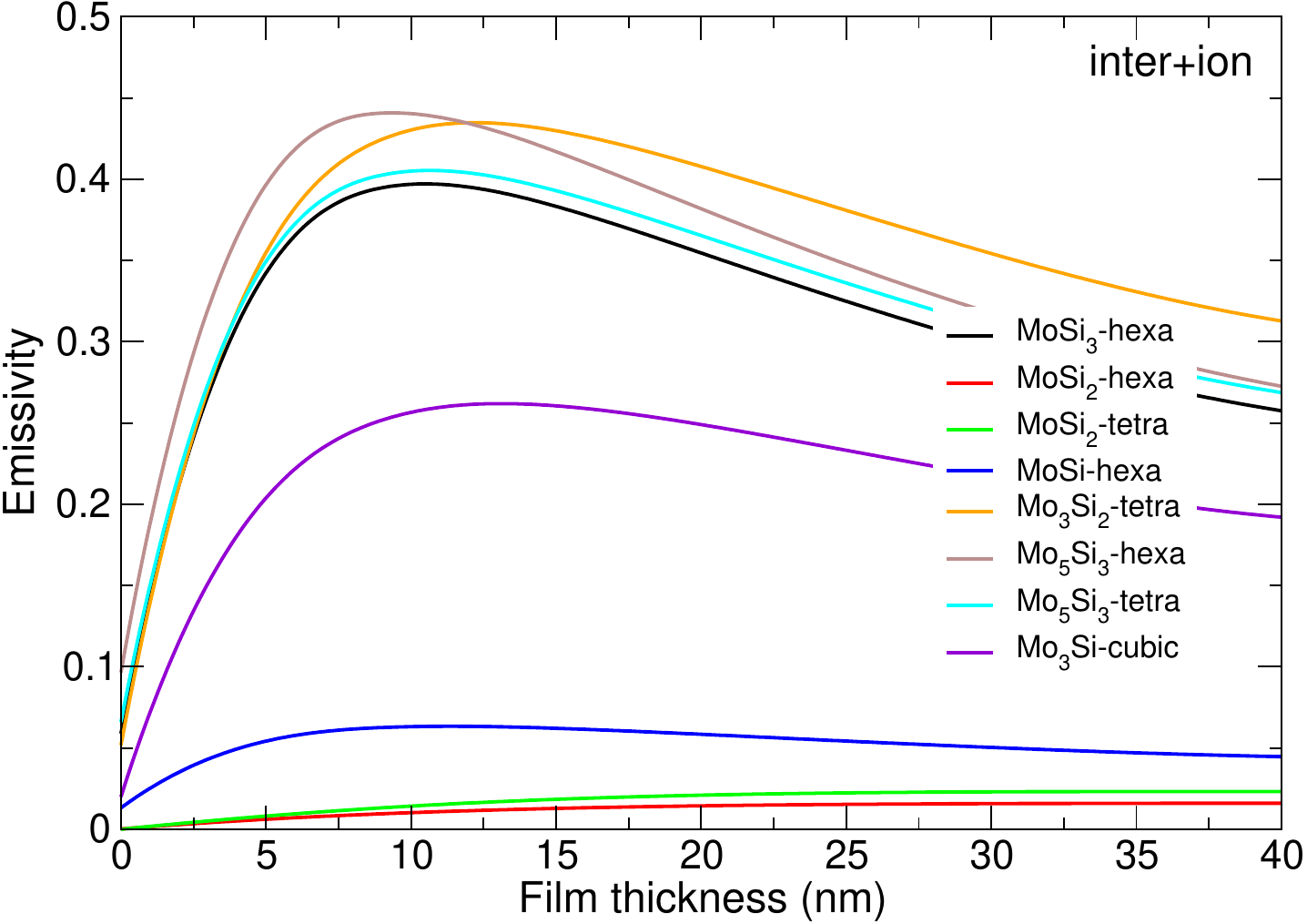}
\includegraphics[width=\columnwidth]{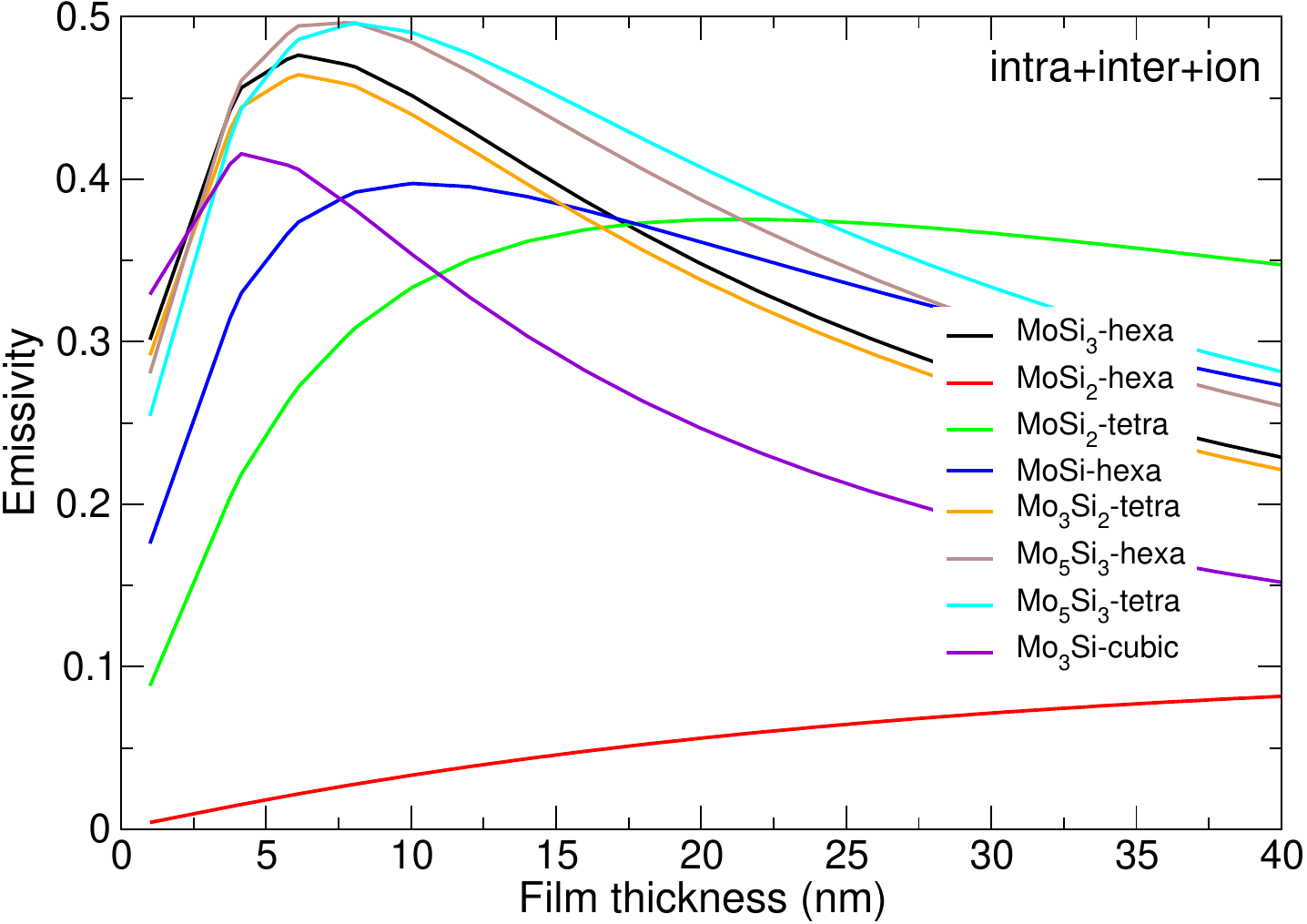}
\caption{Thickness dependence the emissivity of Mo$_x$Si$_y$ thin films at 900~K based on (\textit{top})  inter-band and ionic contributions, and based on (\textit{bottom}) all three contributions.} 
\label{fig11:emis_thick}
\end{figure}

Film thickness is an important parameter for the emissivity of thin films. Figure~\ref{fig11:emis_thick} shows how the emissivity of Mo$_x$Si$_y$ films changes with thickness. Here, the temperature is set to 900~K. As the thickness of the film decreases from 40~nm, the emissivity increases for all Mo$_x$Si$_y$ films, with the exception of hexagonal MoSi$_2$. The emissivity reaches its maximum around $\sim$10~nm. Beyond this point, as the thickness continues to decrease from 10 to 0~nm, the emissivity decreases. Including the intra-band contributions lowers the thickness of the peak maximum to $\sim5$-10~nm depending on the crystal phase. As seen before, MoSi$_2$ stands out from the other films. The difference is related to the presence of a band gap (hexagonal) and low DOS around $\rm E_F$ (tetragonal), resulting in almost no inter-band contributions at low frequency. 

\begin{figure}[t] 
\includegraphics[width=\columnwidth]{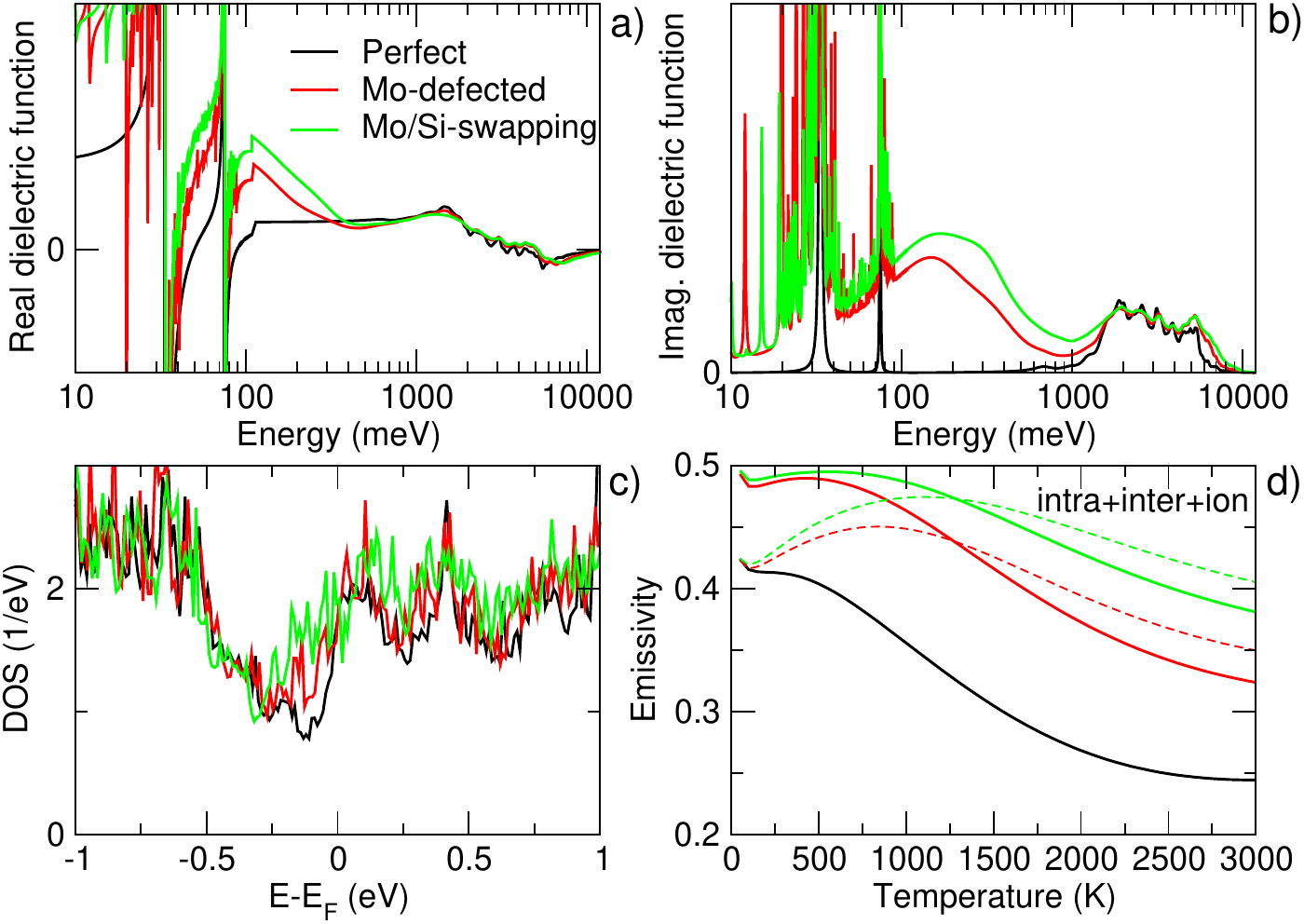}
\caption{Comparison between perfect (black line) and defected (red \&{} green lines) tetragonal MoSi$_2$. a) Real part and b) imaginary part of inter+ion dielectric function, c) density of states, and d) intra+inter+ion emissivity versus temperature. Dashed lines in d) are the calculated emissivities using the same $\omega_{\rm p}$ value for all three systems.} 
\label{fig12:defect}
\end{figure}

\subsection*{Films with reduced crystallinity}
Up to this point, all calculations were based on the crystals depicted in Fig.~\ref{fig:struc}. Conventional techniques to grow Mo$_x$Si$_y$ thin films are based on physical or chemical vapor deposition. Since it is known that these techniques do not produce a perfect crystal structure in the entire film unless special care is taken to control growth (i.e. control of substrate temperature, deposition parameters and precursor materials), even after thermal annealing\cite{BASKAKOV2025}, we study the influence of imperfections in the crystal on the calculated emission coefficient. Here, we select tetragonal-MoSi$_2$ to investigate the influence of defects in the crystal on a qualitative level. This stoichiometry ($\nicefrac{1}{3}$) is most frequently appearing in the literature and, as we have seen, has the peculiar property of low inter-band and high intra-band contribution to the emissivity. As a test cell, we construct a $3\times3\times3$ supercell and consider two kinds of defects: $i$) Mo-\textit{defected}: one additional Mo atom replacing a randomly chosen Si atom in the supercell, and $ii$) Mo\&{}Si-\textit{swapping}: interchanging the position of one Si and one Mo in the supercell. The crystal volume and the atomic coordinates are retained. For the two test cells and for the perfect structure in this supercell, we calculate inter and ion contributions to the dielectric function. The resulting $\varepsilon(\omega)$ are compared to the perfect crystal in Figures~\ref{fig12:defect}~a,b). The defected crystals show a substantial increase of absorption/emission strength in the IR part of the spectrum. This is the result of a stronger response in the energy window between $\sim100$ and 700~meV in the dielectric function. In Fig.~\ref{fig12:defect}~c, we see that the DOS in this frequency range (centered around E$_F$) of the defected systems is larger. Apparently, these states participate in the increase of the amount and dipole-activity of inter-band transitions in the IR regime. The effective ionic contribution to the emissivity is largely unchanged, except for the broadening of the two peaks below 100~meV (see Fig.~\ref{fig12:defect}~b). To study the intra-band contributions, we have calculated the plasma frequencies for the Mo-defected and Mo\&{}Si-swapping test systems as 1.83  and 1.66~eV, respectively. These values are lower than the plasma frequency of the perfect crystal (2.58~eV in supercell). This is to be expected, as the defects break crystal symmetry and add more localized (dispersion-less) electronic states into the band structure. We refer to the SM for the relation between $\omega_{\rm p}$ and the electronic band dispersion. As a result of increased inter-band transitions and decreased $\omega_{\rm p}$ values, the emissivity in both types of defected tetragonal MoSi$_2$ films increases drastically with respect to the perfect crystalline film. Figure~\ref{fig12:defect}~d) shows that this increase is expected to raise the emissivity throughout the here studied temperature range. The dashed lines correspond to the emissivity calculated for the case of same the $\omega_{\rm p}$ value (that of the perfect system) used also for the defected systems. This shows that the emissivity increase at low temperature is mainly attributed to the decrease of the plasma frequency, but the increase around $\sim$1000~K is mainly originating from the increased inter-band transitions. These results indicate that high emissivity for a thin film of MoSi$_2$ most likely means that the atomic structure in the film is defected or partially amorphous.

\section{Discussion}
 
In this computational work, we have mainly studied Mo$_x$Si$_y$ thin films in highly symmetric crystalline form, whereas in experimentally grown thin films there is normally disorder and microstructure. Therefore, the observables reported here should be used as qualitative guidelines and trends, rather than quantitative predictions on the effects of stoichiometric change. For example, the calculated 900~K emissivity for 20~nm thin films of MoSi$_2$ (tetragonal) is 0.37, which is in qualitative agreement with experimental thin film emissivity of 0.33 in Ref.~\cite{BASKAKOV2025}. Also in agreement with experiment, our calculations show a reduced thin film emissivity in the hexagonal crystal phase. We attribute the larger emissivity in the tetragonal as compared to the hexagonal phase, to the closing of the band gap and the resulting sharp increase in the Drude plasma frequency. We note that the calculated thin film emissivity of hexagonal MoSi$_2$ of 0.06 is significantly lower than the experimental value of 0.28. This difference is related to the nanoscale morphology of the film and the presence of defects. 

The emissivity of the thin film can be improved by changing the the Mo$_x$Si$_y$ stoichiometry.  Surprisingly, when the aim is high emissivity films with MoSi$_2$ stoichiometry, we find that one should not focus on the growth of the most perfect crystalline thin film. Rather, our calculations on defected test systems demonstrate increased dipolar activity in the IR part of the dielectric spectrum. We note that during the annealing process of the experimental thin films, the amorphous parts of the film become more crystalline as demonstrated by sharpening of the XRD diffraction spectrum\cite{BASKAKOV2025}. Thereby, annealing of the film probably results in a system with crystalline domains with embedded defects and amorphous parts, i.e. more comparable to the test systems used here. The notable improvement in the emissivity of the defected tetragonal-MoSi$_2$ test system arises in part from an increased contribution of inter-band transitions in the $\sim100$-700~meV window. These calculations only qualitatively demonstrate this effect, a follow up study is needed to determine how representative these test systems are, by calculating the likeliness/energetics of defect formation. We have checked that the effect of increased IR absorption is retained also after structural relaxation of the test systems, be it with somewhat lowered oscillator strength. The corresponding dielectric functions have been added to the SM. We note that this is a truly electronic effect and is not related to a change of thin film morphology. For example, an explanation based on an increase in surface roughness and its effect on the film's emissivity\cite{Simonsen2010} would not fit the model setup chosen here. We furthermore note that, in general, defects do not always increase the emissivity of a semiconductor. Ref.~\cite{Avdoshenko:2014} presents a study of SiO$_2$ and ZrO$_2$ in presence of oxygen vacancies. The study shows that if defects introduce mid-gap states, the emissivity increases (ZrO$_2$), but if it introduces empty states in the conduction bands, the emissivity decreases (SiO$_2$).

Lastly, we would like to discuss two limitations of the applied computational setup. First, the fixed Gaussian smearing level used for all materials results in a small artifact intra-band contribution for MoSi$_2$-hexa. DFT calculations show a small band gap of 0.2 eV, therefore, at 0~K one expects a semiconductor behavior without intra-band contribution, i.e. $\omega_{\rm p}=0$. However, at elevated temperatures, thermal excitations of the electrons can make the material partially metallic. Determining the temperature dependence of the plasma frequency by specific Fermi-Dirac broadening would be an interesting next step. Second, by approximating the dielectric properties of the thin films by their bulk counterpart, the calculated emissivity values for ultra-thin films, ie. below $\sim$5~nm have less predictive value as compared to thicker films. This suggests that the slope of the $\bar{\varepsilon}(T)$ curve in the 0 to 5~nm range of Figure~\ref{fig11:emis_thick} is likely subject to error. In future work, the exact impact of the QSE and of surface reconstruction on the dielectric function can be evaluated through DFPT slab calculations.

\section{Conclusion}
We have calculated the electronic and dielectric properties of different phases of bulk Mo$_x$Si$_y$ crystals using a method based on first-principles. Our findings highlight the significant changes in the dielectric function due to variations in stoichiometry and crystal structure. The calculated frequency-dependent dielectric functions serve as input for further calculations of the optical absorption of thin films. The predicted optical penetration depths are generally larger than the typical thin film thickness of 20~nm, therefore the effect of multiple internal reflections has been included in the calculated absorption. For crystals with high metallicity, such as Mo$_5$Si$_3$, Mo$_3$Si$_2$, Mo$_3$Si, and MoSi$_3$, we find a broad absorption window in the IR window originating from electronic inter- and intra-band contributions. In contrast, compositions such as MoSi$_2$ and MoSi have a much lower inter-band contribution to absorption. This affects their emissivity and, therefore, limit possible high-temperature applications. 

For all eight considered Mo$_x$Si$_y$ materials, we have calculated the emissivity of a thin film as function of temperature. The emissivity shows a complex relation as a function of stoichiometry, which cannot simply be described by the Mo content. In line with experiments of Ref.~\cite{BASKAKOV2025}, we find an increase of the emissivity by changing the crystal phase from hexagonal to tetragonal in a 20~nm thin MoSi$_2$ film at 900~K. The calculated emissivities are in qualitative agreement with these experiments, predicting an emissivity of 0.37 for a $20$~nm tetragonal MoSi$_2$ film at 900~K, compared to $\sim$0.33 in experiment. We demonstrate that ionic contributions have only a small effect on emission at room temperature and become negligible at higher temperatures. Furthermore, our calculations show that atomic defects in MoSi$_2$ introduce states close to the Fermi level, resulting in increased inter-band absorption and a reduction in the intra-band Drude plasma frequency. Together, these effects increase the emissivity of defective MoSi$_2$ relative to perfect tetragonal MoSi$_2$. We have determined that the thickness of the Mo$_x$Si$_y$ film is a crucial factor for high emissivity. The highest emissivity coefficient is predicted for Mo$_5$Si$_3$, Mo$_3$Si$_2$ and MoSi$_3$ thin films with a thickness between 5 and 10 nm. These findings may be useful for the development of applications that require high transmission of short-wave light, by guiding Mo$_x$Si$_y$ growth strategies that lead to thin films with an increased emission coefficient and an optimal film thickness.

\section{Acknowledgment}
This work was jointly funded by the Dutch Ministry of Economic Affairs and ASML within the framework of the TKI-HTSM program, project 2212P21 EMOSION. ZG gratefully acknowledges the support of Fedor Akhmetov for assistance with setting up the TMM absorption calculations. 


\begin{thebibliography}{52}%
\makeatletter
\providecommand \@ifxundefined [1]{%
 \@ifx{#1\undefined}
}%
\providecommand \@ifnum [1]{%
 \ifnum #1\expandafter \@firstoftwo
 \else \expandafter \@secondoftwo
 \fi
}%
\providecommand \@ifx [1]{%
 \ifx #1\expandafter \@firstoftwo
 \else \expandafter \@secondoftwo
 \fi
}%
\providecommand \natexlab [1]{#1}%
\providecommand \enquote  [1]{``#1''}%
\providecommand \bibnamefont  [1]{#1}%
\providecommand \bibfnamefont [1]{#1}%
\providecommand \citenamefont [1]{#1}%
\providecommand \href@noop [0]{\@secondoftwo}%
\providecommand \href [0]{\begingroup \@sanitize@url \@href}%
\providecommand \@href[1]{\@@startlink{#1}\@@href}%
\providecommand \@@href[1]{\endgroup#1\@@endlink}%
\providecommand \@sanitize@url [0]{\catcode `\\12\catcode `\$12\catcode
  `\&12\catcode `\#12\catcode `\^12\catcode `\_12\catcode `\%12\relax}%
\providecommand \@@startlink[1]{}%
\providecommand \@@endlink[0]{}%
\providecommand \url  [0]{\begingroup\@sanitize@url \@url }%
\providecommand \@url [1]{\endgroup\@href {#1}{\urlprefix }}%
\providecommand \urlprefix  [0]{URL }%
\providecommand \Eprint [0]{\href }%
\providecommand \doibase [0]{https://doi.org/}%
\providecommand \selectlanguage [0]{\@gobble}%
\providecommand \bibinfo  [0]{\@secondoftwo}%
\providecommand \bibfield  [0]{\@secondoftwo}%
\providecommand \translation [1]{[#1]}%
\providecommand \BibitemOpen [0]{}%
\providecommand \bibitemStop [0]{}%
\providecommand \bibitemNoStop [0]{.\EOS\space}%
\providecommand \EOS [0]{\spacefactor3000\relax}%
\providecommand \BibitemShut  [1]{\csname bibitem#1\endcsname}%
\let\auto@bib@innerbib\@empty
\bibitem [{\citenamefont {Zhang}\ \emph {et~al.}(2018)\citenamefont {Zhang},
  \citenamefont {Chen},\ and\ \citenamefont {Hu}}]{Zhang:2018}%
  \BibitemOpen
  \bibfield  {author} {\bibinfo {author} {\bibfnamefont {X.}~\bibnamefont
  {Zhang}}, \bibinfo {author} {\bibfnamefont {Y.}~\bibnamefont {Chen}},\ and\
  \bibinfo {author} {\bibfnamefont {J.}~\bibnamefont {Hu}},\ }\bibfield
  {title} {\bibinfo {title} {Recent advances in the development of aerospace
  materials},\ }\href
  {https://doi.org/https://doi.org/10.1016/j.paerosci.2018.01.001} {\bibfield
  {journal} {\bibinfo  {journal} {Progress in Aerospace Sciences}\ }\textbf
  {\bibinfo {volume} {97}},\ \bibinfo {pages} {22} (\bibinfo {year}
  {2018})}\BibitemShut {NoStop}%
\bibitem [{\citenamefont {Werner}\ and\ \citenamefont
  {Fahrner}(2001)}]{werner:2001}%
  \BibitemOpen
  \bibfield  {author} {\bibinfo {author} {\bibfnamefont {M.}~\bibnamefont
  {Werner}}\ and\ \bibinfo {author} {\bibfnamefont {W.}~\bibnamefont
  {Fahrner}},\ }\bibfield  {title} {\bibinfo {title} {Review on materials,
  microsensors, systems and devices for high-temperature and harsh-environment
  applications},\ }\href {https://doi.org/10.1109/41.915402} {\bibfield
  {journal} {\bibinfo  {journal} {IEEE Transactions on Industrial Electronics}\
  }\textbf {\bibinfo {volume} {48}},\ \bibinfo {pages} {249} (\bibinfo {year}
  {2001})}\BibitemShut {NoStop}%
\bibitem [{\citenamefont {Pan}\ \emph {et~al.}(2021)\citenamefont {Pan},
  \citenamefont {Yang}, \citenamefont {Wei}, \citenamefont {Wu}, \citenamefont
  {Dong}, \citenamefont {Wu}, \citenamefont {Wang}, \citenamefont {Zhang},
  \citenamefont {Lin},\ and\ \citenamefont {Mao}}]{PAN:2021}%
  \BibitemOpen
  \bibfield  {author} {\bibinfo {author} {\bibfnamefont {K.}~\bibnamefont
  {Pan}}, \bibinfo {author} {\bibfnamefont {Y.}~\bibnamefont {Yang}}, \bibinfo
  {author} {\bibfnamefont {S.}~\bibnamefont {Wei}}, \bibinfo {author}
  {\bibfnamefont {H.}~\bibnamefont {Wu}}, \bibinfo {author} {\bibfnamefont
  {Z.}~\bibnamefont {Dong}}, \bibinfo {author} {\bibfnamefont {Y.}~\bibnamefont
  {Wu}}, \bibinfo {author} {\bibfnamefont {S.}~\bibnamefont {Wang}}, \bibinfo
  {author} {\bibfnamefont {L.}~\bibnamefont {Zhang}}, \bibinfo {author}
  {\bibfnamefont {J.}~\bibnamefont {Lin}},\ and\ \bibinfo {author}
  {\bibfnamefont {X.}~\bibnamefont {Mao}},\ }\bibfield  {title} {\bibinfo
  {title} {Oxidation behavior of mo-si-b alloys at medium-to-high
  temperatures},\ }\href
  {https://doi.org/https://doi.org/10.1016/j.jmst.2020.06.004} {\bibfield
  {journal} {\bibinfo  {journal} {Journal of Materials Science and Technology}\
  }\textbf {\bibinfo {volume} {60}},\ \bibinfo {pages} {113} (\bibinfo {year}
  {2021})}\BibitemShut {NoStop}%
\bibitem [{\citenamefont {Edalatpour}\ and\ \citenamefont
  {Francoeur}(2013)}]{EDALATPOUR:2013}%
  \BibitemOpen
  \bibfield  {author} {\bibinfo {author} {\bibfnamefont {S.}~\bibnamefont
  {Edalatpour}}\ and\ \bibinfo {author} {\bibfnamefont {M.}~\bibnamefont
  {Francoeur}},\ }\bibfield  {title} {\bibinfo {title} {Size effect on the
  emissivity of thin films},\ }\href
  {https://doi.org/https://doi.org/10.1016/j.jqsrt.2012.12.012} {\bibfield
  {journal} {\bibinfo  {journal} {Journal of Quantitative Spectroscopy and
  Radiative Transfer}\ }\textbf {\bibinfo {volume} {118}},\ \bibinfo {pages}
  {75} (\bibinfo {year} {2013})}\BibitemShut {NoStop}%
\bibitem [{\citenamefont {Baskakov}\ \emph {et~al.}(2025)\citenamefont
  {Baskakov}, \citenamefont {{van de Kruijs}}, \citenamefont {Houweling},
  \citenamefont {Colombi}, \citenamefont {Akhmetov}, \citenamefont {Sturm},\
  and\ \citenamefont {Ackermann}}]{BASKAKOV2025}%
  \BibitemOpen
  \bibfield  {author} {\bibinfo {author} {\bibfnamefont {A.}~\bibnamefont
  {Baskakov}}, \bibinfo {author} {\bibfnamefont {R.}~\bibnamefont {{van de
  Kruijs}}}, \bibinfo {author} {\bibfnamefont {Z.~S.}\ \bibnamefont
  {Houweling}}, \bibinfo {author} {\bibfnamefont {G.}~\bibnamefont {Colombi}},
  \bibinfo {author} {\bibfnamefont {F.}~\bibnamefont {Akhmetov}}, \bibinfo
  {author} {\bibfnamefont {J.~M.}\ \bibnamefont {Sturm}},\ and\ \bibinfo
  {author} {\bibfnamefont {M.}~\bibnamefont {Ackermann}},\ }\bibfield  {title}
  {\bibinfo {title} {Relation between electronic structure and emissivity of
  mosi2-based thin membranes for radiative cooling},\ }\href
  {https://doi.org/https://doi.org/10.1016/j.jallcom.2025.179277} {\bibfield
  {journal} {\bibinfo  {journal} {Journal of Alloys and Compounds}\ }\textbf
  {\bibinfo {volume} {1018}},\ \bibinfo {pages} {179277} (\bibinfo {year}
  {2025})}\BibitemShut {NoStop}%
\bibitem [{\citenamefont {Merchant}\ \emph {et~al.}(2023)\citenamefont
  {Merchant}, \citenamefont {Batzner}, \citenamefont {Schoenholz},
  \citenamefont {Aykol}, \citenamefont {Cheon},\ and\ \citenamefont
  {Cubuk}}]{merchant:2023}%
  \BibitemOpen
  \bibfield  {author} {\bibinfo {author} {\bibfnamefont {A.}~\bibnamefont
  {Merchant}}, \bibinfo {author} {\bibfnamefont {S.}~\bibnamefont {Batzner}},
  \bibinfo {author} {\bibfnamefont {S.~S.}\ \bibnamefont {Schoenholz}},
  \bibinfo {author} {\bibfnamefont {M.}~\bibnamefont {Aykol}}, \bibinfo
  {author} {\bibfnamefont {G.}~\bibnamefont {Cheon}},\ and\ \bibinfo {author}
  {\bibfnamefont {E.~D.}\ \bibnamefont {Cubuk}},\ }\bibfield  {title} {\bibinfo
  {title} {Scaling deep learning for materials discovery},\ }\href
  {https://doi.org/10.1038/s41586-023-06735-9} {\bibfield  {journal} {\bibinfo
  {journal} {Nature}\ }\textbf {\bibinfo {volume} {624}},\ \bibinfo {pages}
  {80} (\bibinfo {year} {2023})}\BibitemShut {NoStop}%
\bibitem [{\citenamefont {Jain}\ \emph {et~al.}(2013)\citenamefont {Jain},
  \citenamefont {Ong}, \citenamefont {Hautier}, \citenamefont {Chen},
  \citenamefont {Richards}, \citenamefont {Dacek}, \citenamefont {Cholia},
  \citenamefont {Gunter}, \citenamefont {Skinner}, \citenamefont {Ceder} \emph
  {et~al.}}]{jain:2013}%
  \BibitemOpen
  \bibfield  {author} {\bibinfo {author} {\bibfnamefont {A.}~\bibnamefont
  {Jain}}, \bibinfo {author} {\bibfnamefont {S.~P.}\ \bibnamefont {Ong}},
  \bibinfo {author} {\bibfnamefont {G.}~\bibnamefont {Hautier}}, \bibinfo
  {author} {\bibfnamefont {W.}~\bibnamefont {Chen}}, \bibinfo {author}
  {\bibfnamefont {W.~D.}\ \bibnamefont {Richards}}, \bibinfo {author}
  {\bibfnamefont {S.}~\bibnamefont {Dacek}}, \bibinfo {author} {\bibfnamefont
  {S.}~\bibnamefont {Cholia}}, \bibinfo {author} {\bibfnamefont
  {D.}~\bibnamefont {Gunter}}, \bibinfo {author} {\bibfnamefont
  {D.}~\bibnamefont {Skinner}}, \bibinfo {author} {\bibfnamefont
  {G.}~\bibnamefont {Ceder}}, \emph {et~al.},\ }\bibfield  {title} {\bibinfo
  {title} {Commentary: The materials project: A materials genome approach to
  accelerating materials innovation},\ }\bibfield  {journal} {\bibinfo
  {journal} {APL materials}\ }\textbf {\bibinfo {volume} {1}},\ \href
  {https://doi.org/10.1063/1.4812323} {10.1063/1.4812323} (\bibinfo {year}
  {2013})\BibitemShut {NoStop}%
\bibitem [{\citenamefont {Brandes}\ and\ \citenamefont
  {Brook}(2013)}]{brandes2013}%
  \BibitemOpen
  \bibfield  {author} {\bibinfo {author} {\bibfnamefont {E.~A.}\ \bibnamefont
  {Brandes}}\ and\ \bibinfo {author} {\bibfnamefont {G.}~\bibnamefont
  {Brook}},\ }\href
  {https://books.google.nl/books?id=lVshBQAAQBAJ&lpg=PP1&ots=0n1fI13Oen&dq=Smithells%20metals%20reference%20book&lr&pg=PP1#v=onepage&q=Smithells%20metals%20reference%20book&f=false}
  {\emph {\bibinfo {title} {Smithells metals reference book}}}\ (\bibinfo
  {publisher} {Elsevier},\ \bibinfo {year} {2013})\BibitemShut {NoStop}%
\bibitem [{\citenamefont {Yao}\ \emph {et~al.}(1999)\citenamefont {Yao},
  \citenamefont {Stiglich},\ and\ \citenamefont {Sudarshan}}]{yao:1999}%
  \BibitemOpen
  \bibfield  {author} {\bibinfo {author} {\bibfnamefont {Z.}~\bibnamefont
  {Yao}}, \bibinfo {author} {\bibfnamefont {J.}~\bibnamefont {Stiglich}},\ and\
  \bibinfo {author} {\bibfnamefont {T.}~\bibnamefont {Sudarshan}},\ }\bibfield
  {title} {\bibinfo {title} {Molybdenum silicide based materials and their
  properties},\ }\href {https://doi.org/10.1361/105994999770346837} {\bibfield
  {journal} {\bibinfo  {journal} {Journal of Materials Engineering and
  Performance}\ }\textbf {\bibinfo {volume} {8}},\ \bibinfo {pages} {291}
  (\bibinfo {year} {1999})}\BibitemShut {NoStop}%
\bibitem [{\citenamefont {Czerny}\ \emph {et~al.}(2024)\citenamefont {Czerny},
  \citenamefont {Ma}, \citenamefont {Hausner}, \citenamefont {Franke},
  \citenamefont {Rohde},\ and\ \citenamefont {Seifert}}]{Czerny:aem24}%
  \BibitemOpen
  \bibfield  {author} {\bibinfo {author} {\bibfnamefont {A.~K.}\ \bibnamefont
  {Czerny}}, \bibinfo {author} {\bibfnamefont {W.}~\bibnamefont {Ma}}, \bibinfo
  {author} {\bibfnamefont {C.~S.}\ \bibnamefont {Hausner}}, \bibinfo {author}
  {\bibfnamefont {P.}~\bibnamefont {Franke}}, \bibinfo {author} {\bibfnamefont
  {M.}~\bibnamefont {Rohde}},\ and\ \bibinfo {author} {\bibfnamefont {H.~J.}\
  \bibnamefont {Seifert}},\ }\bibfield  {title} {\bibinfo {title}
  {Thermodynamic assessment of the mo–si system},\ }\href
  {https://doi.org/https://doi.org/10.1002/adem.202302085} {\bibfield
  {journal} {\bibinfo  {journal} {Advanced Engineering Materials}\ }\textbf
  {\bibinfo {volume} {26}},\ \bibinfo {pages} {2302085} (\bibinfo {year}
  {2024})}\BibitemShut {NoStop}%
\bibitem [{\citenamefont {Vasudévan}\ and\ \citenamefont
  {Petrovic}(1992)}]{VASUDEVAN:1992}%
  \BibitemOpen
  \bibfield  {author} {\bibinfo {author} {\bibfnamefont {A.}~\bibnamefont
  {Vasudévan}}\ and\ \bibinfo {author} {\bibfnamefont {J.}~\bibnamefont
  {Petrovic}},\ }\bibfield  {title} {\bibinfo {title} {A comparative overview
  of molybdenum disilicide composites},\ }\href
  {https://doi.org/https://doi.org/10.1016/0921-5093(92)90308-N} {\bibfield
  {journal} {\bibinfo  {journal} {Materials Science and Engineering: A}\
  }\textbf {\bibinfo {volume} {155}},\ \bibinfo {pages} {1} (\bibinfo {year}
  {1992})},\ \bibinfo {note} {proceedings of the First High Temperature
  Structural Silicides Workshop}\BibitemShut {NoStop}%
\bibitem [{\citenamefont {Dasgupta}\ and\ \citenamefont
  {Umarji}(2008)}]{DASGUPTA:2008}%
  \BibitemOpen
  \bibfield  {author} {\bibinfo {author} {\bibfnamefont {T.}~\bibnamefont
  {Dasgupta}}\ and\ \bibinfo {author} {\bibfnamefont {A.}~\bibnamefont
  {Umarji}},\ }\bibfield  {title} {\bibinfo {title} {Improved ductility and
  oxidation resistance in nb and al co-substituted mosi2},\ }\href
  {https://doi.org/https://doi.org/10.1016/j.intermet.2008.01.006} {\bibfield
  {journal} {\bibinfo  {journal} {Intermetallics}\ }\textbf {\bibinfo {volume}
  {16}},\ \bibinfo {pages} {739} (\bibinfo {year} {2008})}\BibitemShut
  {NoStop}%
\bibitem [{\citenamefont {Bahr}\ \emph {et~al.}(2023)\citenamefont {Bahr},
  \citenamefont {Richter}, \citenamefont {Hahn}, \citenamefont {Wojcik},
  \citenamefont {Podsednik}, \citenamefont {Limbeck}, \citenamefont {Ramm},
  \citenamefont {Hunold}, \citenamefont {Kolozsvári},\ and\ \citenamefont
  {Riedl}}]{BAHR:2023}%
  \BibitemOpen
  \bibfield  {author} {\bibinfo {author} {\bibfnamefont {A.}~\bibnamefont
  {Bahr}}, \bibinfo {author} {\bibfnamefont {S.}~\bibnamefont {Richter}},
  \bibinfo {author} {\bibfnamefont {R.}~\bibnamefont {Hahn}}, \bibinfo {author}
  {\bibfnamefont {T.}~\bibnamefont {Wojcik}}, \bibinfo {author} {\bibfnamefont
  {M.}~\bibnamefont {Podsednik}}, \bibinfo {author} {\bibfnamefont
  {A.}~\bibnamefont {Limbeck}}, \bibinfo {author} {\bibfnamefont
  {J.}~\bibnamefont {Ramm}}, \bibinfo {author} {\bibfnamefont {O.}~\bibnamefont
  {Hunold}}, \bibinfo {author} {\bibfnamefont {S.}~\bibnamefont
  {Kolozsvári}},\ and\ \bibinfo {author} {\bibfnamefont {H.}~\bibnamefont
  {Riedl}},\ }\bibfield  {title} {\bibinfo {title} {Oxidation behaviour and
  mechanical properties of sputter-deposited tmsi2 coatings (tm = mo, ta,
  nb)},\ }\href {https://doi.org/https://doi.org/10.1016/j.jallcom.2022.167532}
  {\bibfield  {journal} {\bibinfo  {journal} {Journal of Alloys and Compounds}\
  }\textbf {\bibinfo {volume} {931}},\ \bibinfo {pages} {167532} (\bibinfo
  {year} {2023})}\BibitemShut {NoStop}%
\bibitem [{\citenamefont {Kumar}\ and\ \citenamefont {Liu}(1993)}]{kumar:1993}%
  \BibitemOpen
  \bibfield  {author} {\bibinfo {author} {\bibfnamefont {K.}~\bibnamefont
  {Kumar}}\ and\ \bibinfo {author} {\bibfnamefont {C.}~\bibnamefont {Liu}},\
  }\bibfield  {title} {\bibinfo {title} {Ordered intermetallic alloys, part ii:
  silicides, trialuminides, and others},\ }\href
  {https://doi.org/10.1007/BF03223307} {\bibfield  {journal} {\bibinfo
  {journal} {JOM}\ }\textbf {\bibinfo {volume} {45}},\ \bibinfo {pages} {28}
  (\bibinfo {year} {1993})}\BibitemShut {NoStop}%
\bibitem [{\citenamefont {Weis}\ \emph {et~al.}(2016)\citenamefont {Weis},
  \citenamefont {Uhlig}, \citenamefont {Wagner}, \citenamefont {Lampke},
  \citenamefont {Bauer},\ and\ \citenamefont {Moldenhauer}}]{weis:2016}%
  \BibitemOpen
  \bibfield  {author} {\bibinfo {author} {\bibfnamefont {S.}~\bibnamefont
  {Weis}}, \bibinfo {author} {\bibfnamefont {T.}~\bibnamefont {Uhlig}},
  \bibinfo {author} {\bibfnamefont {G.}~\bibnamefont {Wagner}}, \bibinfo
  {author} {\bibfnamefont {T.}~\bibnamefont {Lampke}}, \bibinfo {author}
  {\bibfnamefont {W.}~\bibnamefont {Bauer}},\ and\ \bibinfo {author}
  {\bibfnamefont {A.}~\bibnamefont {Moldenhauer}},\ }\bibfield  {title}
  {\bibinfo {title} {High-temperature corrosion and radiation characteristics
  of thermal sprayed molybdenum disilicide-based coatings},\ }\href
  {https://doi.org/10.1088/1757-899X/118/1/012007} {\bibfield  {journal}
  {\bibinfo  {journal} {IOP Conference Series: Materials Science and
  Engineering}\ }\textbf {\bibinfo {volume} {118}},\ \bibinfo {pages} {012007}
  (\bibinfo {year} {2016})}\BibitemShut {NoStop}%
\bibitem [{\citenamefont {Tao}\ \emph {et~al.}(2022)\citenamefont {Tao},
  \citenamefont {Liang}, \citenamefont {Li}, \citenamefont {Zhang},
  \citenamefont {Guo},\ and\ \citenamefont {Wang}}]{tao:2022}%
  \BibitemOpen
  \bibfield  {author} {\bibinfo {author} {\bibfnamefont {X.}~\bibnamefont
  {Tao}}, \bibinfo {author} {\bibfnamefont {Z.}~\bibnamefont {Liang}}, \bibinfo
  {author} {\bibfnamefont {J.}~\bibnamefont {Li}}, \bibinfo {author}
  {\bibfnamefont {J.}~\bibnamefont {Zhang}}, \bibinfo {author} {\bibfnamefont
  {X.}~\bibnamefont {Guo}},\ and\ \bibinfo {author} {\bibfnamefont
  {M.}~\bibnamefont {Wang}},\ }\bibfield  {title} {\bibinfo {title}
  {Anti-oxidation of emissing agents in tasi2–mosi2-borosilicate glass high
  emissivity coating},\ }\href
  {https://doi.org/https://doi.org/10.1016/j.ceramint.2022.08.310} {\bibfield
  {journal} {\bibinfo  {journal} {Ceramics International}\ }\textbf {\bibinfo
  {volume} {48}},\ \bibinfo {pages} {37333} (\bibinfo {year}
  {2022})}\BibitemShut {NoStop}%
\bibitem [{\citenamefont {Yang}\ \emph {et~al.}(2024)\citenamefont {Yang},
  \citenamefont {Wan}, \citenamefont {Li}, \citenamefont {Liu}, \citenamefont
  {Wang},\ and\ \citenamefont {Tao}}]{yang:2023}%
  \BibitemOpen
  \bibfield  {author} {\bibinfo {author} {\bibfnamefont {X.}~\bibnamefont
  {Yang}}, \bibinfo {author} {\bibfnamefont {Y.}~\bibnamefont {Wan}}, \bibinfo
  {author} {\bibfnamefont {J.}~\bibnamefont {Li}}, \bibinfo {author}
  {\bibfnamefont {J.}~\bibnamefont {Liu}}, \bibinfo {author} {\bibfnamefont
  {M.}~\bibnamefont {Wang}},\ and\ \bibinfo {author} {\bibfnamefont
  {X.}~\bibnamefont {Tao}},\ }\bibfield  {title} {\bibinfo {title} {High
  emissivity mosi2-sic-al2o3 coating on rigid insulation tiles with enhanced
  thermal protection performance},\ }\bibfield  {journal} {\bibinfo  {journal}
  {Materials}\ }\textbf {\bibinfo {volume} {17}},\ \href
  {https://doi.org/10.3390/ma17010220} {10.3390/ma17010220} (\bibinfo {year}
  {2024})\BibitemShut {NoStop}%
\bibitem [{\citenamefont {Gokhale}\ and\ \citenamefont
  {Abbaschian}(1991)}]{Gokhale:jope91}%
  \BibitemOpen
  \bibfield  {author} {\bibinfo {author} {\bibfnamefont {A.~B.}\ \bibnamefont
  {Gokhale}}\ and\ \bibinfo {author} {\bibfnamefont {G.~J.}\ \bibnamefont
  {Abbaschian}},\ }\bibfield  {title} {\bibinfo {title} {The mo-si
  (molybdenum-silicon) system},\ }\href {https://doi.org/10.1007/BF02645979}
  {\bibfield  {journal} {\bibinfo  {journal} {Journal of Phase Equilibria}\
  }\textbf {\bibinfo {volume} {12}},\ \bibinfo {pages} {493} (\bibinfo {year}
  {1991})}\BibitemShut {NoStop}%
\bibitem [{\citenamefont {Hawk}\ and\ \citenamefont {Alman}(1995)}]{HAWK:1995}%
  \BibitemOpen
  \bibfield  {author} {\bibinfo {author} {\bibfnamefont {J.}~\bibnamefont
  {Hawk}}\ and\ \bibinfo {author} {\bibfnamefont {D.}~\bibnamefont {Alman}},\
  }\bibfield  {title} {\bibinfo {title} {A comparative study of the abrasive
  wear behavior of mosi2},\ }\href
  {https://doi.org/https://doi.org/10.1016/0956-716X(95)91593-E} {\bibfield
  {journal} {\bibinfo  {journal} {Scripta Metallurgica et Materialia}\ }\textbf
  {\bibinfo {volume} {32}},\ \bibinfo {pages} {725} (\bibinfo {year}
  {1995})}\BibitemShut {NoStop}%
\bibitem [{\citenamefont {Choi}\ \emph {et~al.}(2024)\citenamefont {Choi},
  \citenamefont {Park},\ and\ \citenamefont {Hong}}]{Choi_2024}%
  \BibitemOpen
  \bibfield  {author} {\bibinfo {author} {\bibfnamefont {M.}~\bibnamefont
  {Choi}}, \bibinfo {author} {\bibfnamefont {C.}~\bibnamefont {Park}},\ and\
  \bibinfo {author} {\bibfnamefont {J.}~\bibnamefont {Hong}},\ }\bibfield
  {title} {\bibinfo {title} {Development and optimization of metal silicide euv
  pellicle for 400w euv lithography},\ }\href
  {https://doi.org/10.1088/1361-6528/ad902d} {\bibfield  {journal} {\bibinfo
  {journal} {Nanotechnology}\ }\textbf {\bibinfo {volume} {36}},\ \bibinfo
  {pages} {06LT01} (\bibinfo {year} {2024})}\BibitemShut {NoStop}%
\bibitem [{\citenamefont {Choi}\ and\ \citenamefont {Hong}(2025)}]{Choi_2025}%
  \BibitemOpen
  \bibfield  {author} {\bibinfo {author} {\bibfnamefont {M.}~\bibnamefont
  {Choi}}\ and\ \bibinfo {author} {\bibfnamefont {J.}~\bibnamefont {Hong}},\
  }\bibfield  {title} {\bibinfo {title} {Composition-dependent properties of
  ultra-thin mosix based extreme ultraviolet pellicle},\ }\href
  {https://doi.org/10.1088/1361-6528/add305} {\bibfield  {journal} {\bibinfo
  {journal} {Nanotechnology}\ }\textbf {\bibinfo {volume} {36}},\ \bibinfo
  {pages} {22LT01} (\bibinfo {year} {2025})}\BibitemShut {NoStop}%
\bibitem [{\citenamefont {Liton}\ \emph {et~al.}(2022)\citenamefont {Liton},
  \citenamefont {Helal}, \citenamefont {Khan}, \citenamefont {Kamruzzaman},\
  and\ \citenamefont {Farid Ul~Islam}}]{liton:2022}%
  \BibitemOpen
  \bibfield  {author} {\bibinfo {author} {\bibfnamefont {M.}~\bibnamefont
  {Liton}}, \bibinfo {author} {\bibfnamefont {M.}~\bibnamefont {Helal}},
  \bibinfo {author} {\bibfnamefont {M.}~\bibnamefont {Khan}}, \bibinfo {author}
  {\bibfnamefont {M.}~\bibnamefont {Kamruzzaman}},\ and\ \bibinfo {author}
  {\bibfnamefont {A.}~\bibnamefont {Farid Ul~Islam}},\ }\bibfield  {title}
  {\bibinfo {title} {Mechanical and opto-electronic properties of
  $\alpha$-mosi2: a dft scheme with hydrostatic pressure},\ }\href
  {https://doi.org/10.1007/s12648-022-02355-7} {\bibfield  {journal} {\bibinfo
  {journal} {Indian Journal of Physics}\ }\textbf {\bibinfo {volume} {96}},\
  \bibinfo {pages} {4155} (\bibinfo {year} {2022})}\BibitemShut {NoStop}%
\bibitem [{\citenamefont {Zheng}\ \emph {et~al.}(2019)\citenamefont {Zheng},
  \citenamefont {Chen}, \citenamefont {Zhang}, \citenamefont {Liu},
  \citenamefont {Chen}, \citenamefont {Wu},\ and\ \citenamefont
  {Huang}}]{Zheng:2019}%
  \BibitemOpen
  \bibfield  {author} {\bibinfo {author} {\bibfnamefont {J.-Q.}\ \bibnamefont
  {Zheng}}, \bibinfo {author} {\bibfnamefont {J.}~\bibnamefont {Chen}},
  \bibinfo {author} {\bibfnamefont {B.-H.}\ \bibnamefont {Zhang}}, \bibinfo
  {author} {\bibfnamefont {X.-J.}\ \bibnamefont {Liu}}, \bibinfo {author}
  {\bibfnamefont {Z.-M.}\ \bibnamefont {Chen}}, \bibinfo {author}
  {\bibfnamefont {H.-B.}\ \bibnamefont {Wu}},\ and\ \bibinfo {author}
  {\bibfnamefont {Z.-R.}\ \bibnamefont {Huang}},\ }\bibfield  {title} {\bibinfo
  {title} {Electrical percolation and infrared emissivity of pressureless
  sintered sic-mosi2 composites tailored by sintering temperature},\ }\href
  {https://doi.org/https://doi.org/10.1016/j.jeurceramsoc.2019.05.019}
  {\bibfield  {journal} {\bibinfo  {journal} {Journal of the European Ceramic
  Society}\ }\textbf {\bibinfo {volume} {39}},\ \bibinfo {pages} {3981}
  (\bibinfo {year} {2019})}\BibitemShut {NoStop}%
\bibitem [{\citenamefont {van Setten}\ \emph {et~al.}(2009)\citenamefont {van
  Setten}, \citenamefont {Er}, \citenamefont {Brocks}, \citenamefont
  {de~Groot},\ and\ \citenamefont {de~Wijs}}]{vanZetten:prb09}%
  \BibitemOpen
  \bibfield  {author} {\bibinfo {author} {\bibfnamefont {M.~J.}\ \bibnamefont
  {van Setten}}, \bibinfo {author} {\bibfnamefont {S.}~\bibnamefont {Er}},
  \bibinfo {author} {\bibfnamefont {G.}~\bibnamefont {Brocks}}, \bibinfo
  {author} {\bibfnamefont {R.~A.}\ \bibnamefont {de~Groot}},\ and\ \bibinfo
  {author} {\bibfnamefont {G.~A.}\ \bibnamefont {de~Wijs}},\ }\bibfield
  {title} {\bibinfo {title} {First-principles study of the optical properties
  of ${\text{mg}}_{x}{\text{ti}}_{1\ensuremath{-}x}{\text{h}}_{2}$},\ }\href
  {https://doi.org/10.1103/PhysRevB.79.125117} {\bibfield  {journal} {\bibinfo
  {journal} {Phys. Rev. B}\ }\textbf {\bibinfo {volume} {79}},\ \bibinfo
  {pages} {125117} (\bibinfo {year} {2009})}\BibitemShut {NoStop}%
\bibitem [{\citenamefont {Jiao}\ \emph {et~al.}(2011)\citenamefont {Jiao},
  \citenamefont {Ma},\ and\ \citenamefont {Yang}}]{jiao:2011}%
  \BibitemOpen
  \bibfield  {author} {\bibinfo {author} {\bibfnamefont {Z.-Y.}\ \bibnamefont
  {Jiao}}, \bibinfo {author} {\bibfnamefont {S.-H.}\ \bibnamefont {Ma}},\ and\
  \bibinfo {author} {\bibfnamefont {J.-F.}\ \bibnamefont {Yang}},\ }\bibfield
  {title} {\bibinfo {title} {A comparison of the electronic and optical
  properties of zinc-blende, rocksalt and wurtzite aln: A dft study},\ }\href
  {https://doi.org/https://doi.org/10.1016/j.solidstatesciences.2010.11.030}
  {\bibfield  {journal} {\bibinfo  {journal} {Solid State Sciences}\ }\textbf
  {\bibinfo {volume} {13}},\ \bibinfo {pages} {331} (\bibinfo {year}
  {2011})}\BibitemShut {NoStop}%
\bibitem [{\citenamefont {Ming}\ \emph {et~al.}(2014)\citenamefont {Ming},
  \citenamefont {Blair},\ and\ \citenamefont {Liu}}]{ming:2014}%
  \BibitemOpen
  \bibfield  {author} {\bibinfo {author} {\bibfnamefont {W.}~\bibnamefont
  {Ming}}, \bibinfo {author} {\bibfnamefont {S.}~\bibnamefont {Blair}},\ and\
  \bibinfo {author} {\bibfnamefont {F.}~\bibnamefont {Liu}},\ }\bibfield
  {title} {\bibinfo {title} {Quantum size effect on dielectric function of
  ultrathin metal film: a first-principles study of al 111},\ }\href
  {https://doi.org/10.1088/0953-8984/26/50/505302} {\bibfield  {journal}
  {\bibinfo  {journal} {Journal of Physics: Condensed Matter}\ }\textbf
  {\bibinfo {volume} {26}},\ \bibinfo {pages} {505302} (\bibinfo {year}
  {2014})}\BibitemShut {NoStop}%
\bibitem [{\citenamefont {Xiang}\ \emph {et~al.}(2021)\citenamefont {Xiang},
  \citenamefont {Dai},\ and\ \citenamefont {Zhou}}]{XIANG:2021}%
  \BibitemOpen
  \bibfield  {author} {\bibinfo {author} {\bibfnamefont {H.}~\bibnamefont
  {Xiang}}, \bibinfo {author} {\bibfnamefont {F.}~\bibnamefont {Dai}},\ and\
  \bibinfo {author} {\bibfnamefont {Y.}~\bibnamefont {Zhou}},\ }\bibfield
  {title} {\bibinfo {title} {Secrets of high thermal emission of transition
  metal disilicides tmsi2 (tm=ta, mo)},\ }\href
  {https://doi.org/https://doi.org/10.1016/j.jmst.2021.02.026} {\bibfield
  {journal} {\bibinfo  {journal} {Journal of Materials Science and Technology}\
  }\textbf {\bibinfo {volume} {89}},\ \bibinfo {pages} {114} (\bibinfo {year}
  {2021})}\BibitemShut {NoStop}%
\bibitem [{\citenamefont {Song}\ \emph {et~al.}(2023)\citenamefont {Song},
  \citenamefont {Cheng}, \citenamefont {Xiang}, \citenamefont {Dai},
  \citenamefont {Dong}, \citenamefont {Chen}, \citenamefont {Hu}, \citenamefont
  {Zhang}, \citenamefont {Han},\ and\ \citenamefont {Zhou}}]{SONG2023}%
  \BibitemOpen
  \bibfield  {author} {\bibinfo {author} {\bibfnamefont {J.}~\bibnamefont
  {Song}}, \bibinfo {author} {\bibfnamefont {Y.}~\bibnamefont {Cheng}},
  \bibinfo {author} {\bibfnamefont {H.}~\bibnamefont {Xiang}}, \bibinfo
  {author} {\bibfnamefont {F.-Z.}\ \bibnamefont {Dai}}, \bibinfo {author}
  {\bibfnamefont {S.}~\bibnamefont {Dong}}, \bibinfo {author} {\bibfnamefont
  {G.}~\bibnamefont {Chen}}, \bibinfo {author} {\bibfnamefont {P.}~\bibnamefont
  {Hu}}, \bibinfo {author} {\bibfnamefont {X.}~\bibnamefont {Zhang}}, \bibinfo
  {author} {\bibfnamefont {W.}~\bibnamefont {Han}},\ and\ \bibinfo {author}
  {\bibfnamefont {Y.}~\bibnamefont {Zhou}},\ }\bibfield  {title} {\bibinfo
  {title} {Medium and high-entropy transition mental disilicides with improved
  infrared emissivity for thermal protection applications},\ }\href
  {https://doi.org/10.1016/j.jmst.2022.07.028} {\bibfield  {journal} {\bibinfo
  {journal} {J. Mat. Sci. \& Tech.}\ }\textbf {\bibinfo {volume} {136}},\
  \bibinfo {pages} {149} (\bibinfo {year} {2023})}\BibitemShut {NoStop}%
\bibitem [{\citenamefont {Bl\"{o}chl}(1994)}]{paw-1}%
  \BibitemOpen
  \bibfield  {author} {\bibinfo {author} {\bibfnamefont {P.~E.}\ \bibnamefont
  {Bl\"{o}chl}},\ }\bibfield  {title} {\bibinfo {title} {Projector
  augmented-wave method},\ }\href {https://doi.org/10.1103/PhysRevB.50.17953}
  {\bibfield  {journal} {\bibinfo  {journal} {Phys. Rev. B}\ }\textbf {\bibinfo
  {volume} {50}},\ \bibinfo {pages} {17953} (\bibinfo {year}
  {1994})}\BibitemShut {NoStop}%
\bibitem [{\citenamefont {Kresse}\ and\ \citenamefont
  {Hafner}(1993)}]{Kresse:prb93}%
  \BibitemOpen
  \bibfield  {author} {\bibinfo {author} {\bibfnamefont {G.}~\bibnamefont
  {Kresse}}\ and\ \bibinfo {author} {\bibfnamefont {J.}~\bibnamefont
  {Hafner}},\ }\bibfield  {title} {\bibinfo {title} {Ab-initio
  molecular-dynamics for liquid-metals},\ }\href
  {https://doi.org/10.1103/PhysRevB.47.558} {\bibfield  {journal} {\bibinfo
  {journal} {Phys. Rev. B}\ }\textbf {\bibinfo {volume} {47}},\ \bibinfo
  {pages} {558} (\bibinfo {year} {1993})}\BibitemShut {NoStop}%
\bibitem [{\citenamefont {Kresse}\ and\ \citenamefont
  {Furthm{\"{u}}ller}(1996)}]{Kresse:prb96}%
  \BibitemOpen
  \bibfield  {author} {\bibinfo {author} {\bibfnamefont {G.}~\bibnamefont
  {Kresse}}\ and\ \bibinfo {author} {\bibfnamefont {J.}~\bibnamefont
  {Furthm{\"{u}}ller}},\ }\bibfield  {title} {\bibinfo {title} {Efficient
  iterative schemes for ab initio total-energy calculations using a plane-wave
  basis set},\ }\href {https://doi.org/10.1103/PhysRevB.54.11169} {\bibfield
  {journal} {\bibinfo  {journal} {Phys. Rev. B}\ }\textbf {\bibinfo {volume}
  {54}},\ \bibinfo {pages} {11169} (\bibinfo {year} {1996})}\BibitemShut
  {NoStop}%
\bibitem [{\citenamefont {Kresse}\ and\ \citenamefont {Joubert}(1999)}]{paw-2}%
  \BibitemOpen
  \bibfield  {author} {\bibinfo {author} {\bibfnamefont {G.}~\bibnamefont
  {Kresse}}\ and\ \bibinfo {author} {\bibfnamefont {D.}~\bibnamefont
  {Joubert}},\ }\bibfield  {title} {\bibinfo {title} {From ultrasoft
  pseudopotentials to the projector augmented-wave method},\ }\href
  {https://doi.org/10.1103/PhysRevB.59.1758} {\bibfield  {journal} {\bibinfo
  {journal} {Phys. Rev. B}\ }\textbf {\bibinfo {volume} {59}},\ \bibinfo
  {pages} {1758} (\bibinfo {year} {1999})}\BibitemShut {NoStop}%
\bibitem [{\citenamefont {Perdew}\ \emph {et~al.}(2008)\citenamefont {Perdew},
  \citenamefont {Ruzsinszky}, \citenamefont {Csonka}, \citenamefont {Vydrov},
  \citenamefont {Scuseria}, \citenamefont {Constantin}, \citenamefont {Zhou},\
  and\ \citenamefont {Burke}}]{perdew:prl2008}%
  \BibitemOpen
  \bibfield  {author} {\bibinfo {author} {\bibfnamefont {J.~P.}\ \bibnamefont
  {Perdew}}, \bibinfo {author} {\bibfnamefont {A.}~\bibnamefont {Ruzsinszky}},
  \bibinfo {author} {\bibfnamefont {G.~I.}\ \bibnamefont {Csonka}}, \bibinfo
  {author} {\bibfnamefont {O.~A.}\ \bibnamefont {Vydrov}}, \bibinfo {author}
  {\bibfnamefont {G.~E.}\ \bibnamefont {Scuseria}}, \bibinfo {author}
  {\bibfnamefont {L.~A.}\ \bibnamefont {Constantin}}, \bibinfo {author}
  {\bibfnamefont {X.}~\bibnamefont {Zhou}},\ and\ \bibinfo {author}
  {\bibfnamefont {K.}~\bibnamefont {Burke}},\ }\bibfield  {title} {\bibinfo
  {title} {Restoring the density-gradient expansion for exchange in solids and
  surfaces},\ }\href {https://doi.org/10.1103/PhysRevLett.100.136406}
  {\bibfield  {journal} {\bibinfo  {journal} {Phys. Rev. Lett.}\ }\textbf
  {\bibinfo {volume} {100}},\ \bibinfo {pages} {136406} (\bibinfo {year}
  {2008})}\BibitemShut {NoStop}%
\bibitem [{\citenamefont {Gonze}\ and\ \citenamefont
  {Lee}(1997)}]{Gonze:prb97}%
  \BibitemOpen
  \bibfield  {author} {\bibinfo {author} {\bibfnamefont {X.}~\bibnamefont
  {Gonze}}\ and\ \bibinfo {author} {\bibfnamefont {C.}~\bibnamefont {Lee}},\
  }\bibfield  {title} {\bibinfo {title} {Dynamical matrices, born effective
  charges, dielectric permittivity tensors, and interatomic force constants
  from density-functional perturbation theory},\ }\href
  {https://doi.org/10.1103/PhysRevB.55.10355} {\bibfield  {journal} {\bibinfo
  {journal} {Phys. Rev. B}\ }\textbf {\bibinfo {volume} {55}},\ \bibinfo
  {pages} {10355} (\bibinfo {year} {1997})}\BibitemShut {NoStop}%
\bibitem [{\citenamefont {Wu}\ \emph {et~al.}(2005)\citenamefont {Wu},
  \citenamefont {Vanderbilt},\ and\ \citenamefont {Hamann}}]{Wu:prb05}%
  \BibitemOpen
  \bibfield  {author} {\bibinfo {author} {\bibfnamefont {X.}~\bibnamefont
  {Wu}}, \bibinfo {author} {\bibfnamefont {D.}~\bibnamefont {Vanderbilt}},\
  and\ \bibinfo {author} {\bibfnamefont {D.~R.}\ \bibnamefont {Hamann}},\
  }\bibfield  {title} {\bibinfo {title} {Systematic treatment of displacements,
  strains, and electric fields in density-functional perturbation theory},\
  }\href {https://doi.org/10.1103/PhysRevB.72.035105} {\bibfield  {journal}
  {\bibinfo  {journal} {Phys. Rev. B}\ }\textbf {\bibinfo {volume} {72}},\
  \bibinfo {pages} {035105} (\bibinfo {year} {2005})}\BibitemShut {NoStop}%
\bibitem [{\citenamefont {Gajdo\ifmmode~\check{s}\else \v{s}\fi{}}\ \emph
  {et~al.}(2006)\citenamefont {Gajdo\ifmmode~\check{s}\else \v{s}\fi{}},
  \citenamefont {Hummer}, \citenamefont {Kresse}, \citenamefont
  {Furthm\"uller},\ and\ \citenamefont {Bechstedt}}]{Gajdos:prb06}%
  \BibitemOpen
  \bibfield  {author} {\bibinfo {author} {\bibfnamefont {M.}~\bibnamefont
  {Gajdo\ifmmode~\check{s}\else \v{s}\fi{}}}, \bibinfo {author} {\bibfnamefont
  {K.}~\bibnamefont {Hummer}}, \bibinfo {author} {\bibfnamefont
  {G.}~\bibnamefont {Kresse}}, \bibinfo {author} {\bibfnamefont
  {J.}~\bibnamefont {Furthm\"uller}},\ and\ \bibinfo {author} {\bibfnamefont
  {F.}~\bibnamefont {Bechstedt}},\ }\bibfield  {title} {\bibinfo {title}
  {Linear optical properties in the projector-augmented wave methodology},\
  }\href {https://doi.org/10.1103/PhysRevB.73.045112} {\bibfield  {journal}
  {\bibinfo  {journal} {Phys. Rev. B}\ }\textbf {\bibinfo {volume} {73}},\
  \bibinfo {pages} {045112} (\bibinfo {year} {2006})}\BibitemShut {NoStop}%
\bibitem [{\citenamefont {Lee}\ and\ \citenamefont
  {Chang}(1994)}]{LeeKeun-HoChang:prb94}%
  \BibitemOpen
  \bibfield  {author} {\bibinfo {author} {\bibfnamefont {K.-H.}\ \bibnamefont
  {Lee}}\ and\ \bibinfo {author} {\bibfnamefont {K.~J.}\ \bibnamefont
  {Chang}},\ }\bibfield  {title} {\bibinfo {title} {First-principles study of
  the optical properties and the dielectric response of al},\ }\href
  {https://doi.org/10.1103/PhysRevB.49.2362} {\bibfield  {journal} {\bibinfo
  {journal} {Phys. Rev. B}\ }\textbf {\bibinfo {volume} {49}},\ \bibinfo
  {pages} {2362} (\bibinfo {year} {1994})}\BibitemShut {NoStop}%
\bibitem [{\citenamefont {Harl}\ \emph {et~al.}(2007)\citenamefont {Harl},
  \citenamefont {Kresse}, \citenamefont {Sun}, \citenamefont {Hohage},\ and\
  \citenamefont {Zeppenfeld}}]{Harl:prb07}%
  \BibitemOpen
  \bibfield  {author} {\bibinfo {author} {\bibfnamefont {J.}~\bibnamefont
  {Harl}}, \bibinfo {author} {\bibfnamefont {G.}~\bibnamefont {Kresse}},
  \bibinfo {author} {\bibfnamefont {L.~D.}\ \bibnamefont {Sun}}, \bibinfo
  {author} {\bibfnamefont {M.}~\bibnamefont {Hohage}},\ and\ \bibinfo {author}
  {\bibfnamefont {P.}~\bibnamefont {Zeppenfeld}},\ }\bibfield  {title}
  {\bibinfo {title} {Ab initio reflectance difference spectra of the bare and
  adsorbate covered cu(110) surfaces},\ }\href
  {https://doi.org/10.1103/PhysRevB.76.035436} {\bibfield  {journal} {\bibinfo
  {journal} {Phys. Rev. B}\ }\textbf {\bibinfo {volume} {76}},\ \bibinfo
  {pages} {035436} (\bibinfo {year} {2007})}\BibitemShut {NoStop}%
\bibitem [{sup()}]{supp}%
  \BibitemOpen
  \href@noop {} {}\bibinfo {note} {See Supplemental Material at \href{http://link.aps.org/supplemental/10.1103/6v8t-117b}{http://link.aps.org/supplemental/10.1103/6v8t-117b} for
  supporting information about the calculations, which includes Ref.
  \cite{Harl08}.}\BibitemShut {Stop}%
\bibitem [{\citenamefont {Bokdam}\ \emph {et~al.}(2016)\citenamefont {Bokdam},
  \citenamefont {Sander}, \citenamefont {Stroppa}, \citenamefont {Picozzi},
  \citenamefont {Sarma}, \citenamefont {Franchini},\ and\ \citenamefont
  {Kresse}}]{Bokdam:sr16}%
  \BibitemOpen
  \bibfield  {author} {\bibinfo {author} {\bibfnamefont {M.}~\bibnamefont
  {Bokdam}}, \bibinfo {author} {\bibfnamefont {T.}~\bibnamefont {Sander}},
  \bibinfo {author} {\bibfnamefont {A.}~\bibnamefont {Stroppa}}, \bibinfo
  {author} {\bibfnamefont {S.}~\bibnamefont {Picozzi}}, \bibinfo {author}
  {\bibfnamefont {D.~D.}\ \bibnamefont {Sarma}}, \bibinfo {author}
  {\bibfnamefont {C.}~\bibnamefont {Franchini}},\ and\ \bibinfo {author}
  {\bibfnamefont {G.}~\bibnamefont {Kresse}},\ }\bibfield  {title} {\bibinfo
  {title} {Role of polar phonons in the photo excited state of metal halide
  perovskites},\ }\href {https://doi.org/10.1038/srep28618} {\bibfield
  {journal} {\bibinfo  {journal} {Sci. Rep.}\ }\textbf {\bibinfo {volume}
  {6}},\ \bibinfo {pages} {28618} (\bibinfo {year} {2016})}\BibitemShut
  {NoStop}%
\bibitem [{\citenamefont {Zubizarreta}\ \emph {et~al.}(2017)\citenamefont
  {Zubizarreta}, \citenamefont {Chulkov}, \citenamefont {Chernov},
  \citenamefont {Vasenko}, \citenamefont {Aldazabal},\ and\ \citenamefont
  {Silkin}}]{Zubizarreta:prb17}%
  \BibitemOpen
  \bibfield  {author} {\bibinfo {author} {\bibfnamefont {X.}~\bibnamefont
  {Zubizarreta}}, \bibinfo {author} {\bibfnamefont {E.~V.}\ \bibnamefont
  {Chulkov}}, \bibinfo {author} {\bibfnamefont {I.~P.}\ \bibnamefont
  {Chernov}}, \bibinfo {author} {\bibfnamefont {A.~S.}\ \bibnamefont
  {Vasenko}}, \bibinfo {author} {\bibfnamefont {I.}~\bibnamefont {Aldazabal}},\
  and\ \bibinfo {author} {\bibfnamefont {V.~M.}\ \bibnamefont {Silkin}},\
  }\bibfield  {title} {\bibinfo {title} {Quantum-size effects in the loss
  function of pb(111) thin films: An ab initio study},\ }\href
  {https://doi.org/10.1103/PhysRevB.95.235405} {\bibfield  {journal} {\bibinfo
  {journal} {Phys. Rev. B}\ }\textbf {\bibinfo {volume} {95}},\ \bibinfo
  {pages} {235405} (\bibinfo {year} {2017})}\BibitemShut {NoStop}%
\bibitem [{\citenamefont {Byrnes}(2020)}]{byrnes:arxive2020}%
  \BibitemOpen
  \bibfield  {author} {\bibinfo {author} {\bibfnamefont {S.~J.}\ \bibnamefont
  {Byrnes}},\ }\href@noop {} {\bibinfo {title} {Multilayer optical
  calculations}} (\bibinfo {year} {2020}),\ \Eprint
  {https://arxiv.org/abs/1603.02720} {arXiv:1603.02720 [physics.comp-ph]}
  \BibitemShut {NoStop}%
\bibitem [{TMM()}]{TMMabs}%
  \BibitemOpen
  \href@noop {} {\bibinfo {title} {Absorptiontmm}},\ \bibinfo {howpublished}
  {\url{https://github.com/udcm-su/AbsorptionTMM}},\ \bibinfo {note} {accessed:
  2024-08-10}\BibitemShut {NoStop}%
\bibitem [{ost(2020)}]{osti_1315941}%
  \BibitemOpen
  \bibfield  {title} {\bibinfo {title} {Materials data on si3mo by materials
  project}\ }\href {https://doi.org/10.17188/1315941} {10.17188/1315941}
  (\bibinfo {year} {2020})\BibitemShut {NoStop}%
\bibitem [{\citenamefont {Woods-Robinson}\ \emph {et~al.}(2023)\citenamefont
  {Woods-Robinson}, \citenamefont {Xiong}, \citenamefont {Shen}, \citenamefont
  {Winner}, \citenamefont {Horton}, \citenamefont {Asta}, \citenamefont
  {Ganose}, \citenamefont {Hautier},\ and\ \citenamefont
  {Persson}}]{Woods:matter23}%
  \BibitemOpen
  \bibfield  {author} {\bibinfo {author} {\bibfnamefont {R.}~\bibnamefont
  {Woods-Robinson}}, \bibinfo {author} {\bibfnamefont {Y.}~\bibnamefont
  {Xiong}}, \bibinfo {author} {\bibfnamefont {J.-X.}\ \bibnamefont {Shen}},
  \bibinfo {author} {\bibfnamefont {N.}~\bibnamefont {Winner}}, \bibinfo
  {author} {\bibfnamefont {M.~K.}\ \bibnamefont {Horton}}, \bibinfo {author}
  {\bibfnamefont {M.}~\bibnamefont {Asta}}, \bibinfo {author} {\bibfnamefont
  {A.~M.}\ \bibnamefont {Ganose}}, \bibinfo {author} {\bibfnamefont
  {G.}~\bibnamefont {Hautier}},\ and\ \bibinfo {author} {\bibfnamefont {K.~A.}\
  \bibnamefont {Persson}},\ }\bibfield  {title} {\bibinfo {title} {Designing
  transparent conductors using forbidden optical transitions},\ }\href
  {https://doi.org/10.1016/j.matt.2023.06.043} {\bibfield  {journal} {\bibinfo
  {journal} {Matter}\ }\textbf {\bibinfo {volume} {6}},\ \bibinfo {pages}
  {3021} (\bibinfo {year} {2023})}\BibitemShut {NoStop}%
\bibitem [{\citenamefont {Setyawan}\ and\ \citenamefont
  {Curtarolo}(2010)}]{Setyawan:ComMatSci2010}%
  \BibitemOpen
  \bibfield  {author} {\bibinfo {author} {\bibfnamefont {W.}~\bibnamefont
  {Setyawan}}\ and\ \bibinfo {author} {\bibfnamefont {S.}~\bibnamefont
  {Curtarolo}},\ }\bibfield  {title} {\bibinfo {title} {High-throughput
  electronic band structure calculations: Challenges and tools},\ }\href
  {https://doi.org/https://doi.org/10.1016/j.commatsci.2010.05.010} {\bibfield
  {journal} {\bibinfo  {journal} {Computational Materials Science}\ }\textbf
  {\bibinfo {volume} {49}},\ \bibinfo {pages} {299} (\bibinfo {year}
  {2010})}\BibitemShut {NoStop}%
\bibitem [{\citenamefont {Shah}\ \emph {et~al.}(2022)\citenamefont {Shah},
  \citenamefont {Yang}, \citenamefont {Kudyshev}, \citenamefont {Xu},
  \citenamefont {Shalaev}, \citenamefont {Bondarev},\ and\ \citenamefont
  {Boltasseva}}]{Shah2022}%
  \BibitemOpen
  \bibfield  {author} {\bibinfo {author} {\bibfnamefont {D.}~\bibnamefont
  {Shah}}, \bibinfo {author} {\bibfnamefont {M.}~\bibnamefont {Yang}}, \bibinfo
  {author} {\bibfnamefont {Z.}~\bibnamefont {Kudyshev}}, \bibinfo {author}
  {\bibfnamefont {X.}~\bibnamefont {Xu}}, \bibinfo {author} {\bibfnamefont
  {V.~M.}\ \bibnamefont {Shalaev}}, \bibinfo {author} {\bibfnamefont {I.~V.}\
  \bibnamefont {Bondarev}},\ and\ \bibinfo {author} {\bibfnamefont
  {A.}~\bibnamefont {Boltasseva}},\ }\bibfield  {title} {\bibinfo {title}
  {Thickness-dependent drude plasma frequency in transdimensional plasmonic
  tin},\ }\href {https://doi.org/10.1021/acs.nanolett.1c04692} {\bibfield
  {journal} {\bibinfo  {journal} {Nano Lett.}\ }\textbf {\bibinfo {volume}
  {22}},\ \bibinfo {pages} {4622} (\bibinfo {year} {2022})}\BibitemShut
  {NoStop}%
\bibitem [{\citenamefont {Siegel}(2001)}]{Siegel:book01}%
  \BibitemOpen
  \bibfield  {author} {\bibinfo {author} {\bibfnamefont {R.}~\bibnamefont
  {Siegel}},\ }\href@noop {} {\emph {\bibinfo {title} {Thermal Radiation Heat
  Transfer}}}\ (\bibinfo  {publisher} {CRC Press},\ \bibinfo {year}
  {2001})\BibitemShut {NoStop}%
\bibitem [{\citenamefont {Hadley}\ and\ \citenamefont
  {Dennison}(1947)}]{Hadley:47}%
  \BibitemOpen
  \bibfield  {author} {\bibinfo {author} {\bibfnamefont {L.~N.}\ \bibnamefont
  {Hadley}}\ and\ \bibinfo {author} {\bibfnamefont {D.~M.}\ \bibnamefont
  {Dennison}},\ }\bibfield  {title} {\bibinfo {title} {Reflection and
  transmission interference filters part i. theory},\ }\href
  {https://doi.org/10.1364/JOSA.37.000451} {\bibfield  {journal} {\bibinfo
  {journal} {J. Opt. Soc. Am.}\ }\textbf {\bibinfo {volume} {37}},\ \bibinfo
  {pages} {451} (\bibinfo {year} {1947})}\BibitemShut {NoStop}%
\bibitem [{\citenamefont {Simonsen}(2010)}]{Simonsen2010}%
  \BibitemOpen
  \bibfield  {author} {\bibinfo {author} {\bibfnamefont {I.}~\bibnamefont
  {Simonsen}},\ }\bibfield  {title} {\bibinfo {title} {Optics of surface
  disordered systems},\ }\href {https://doi.org/10.1140/epjst/e2010-01221-4}
  {\bibfield  {journal} {\bibinfo  {journal} {The European Physical Journal
  Special Topics}\ }\textbf {\bibinfo {volume} {181}},\ \bibinfo {pages} {1}
  (\bibinfo {year} {2010})}\BibitemShut {NoStop}%
\bibitem [{\citenamefont {Avdoshenko}\ and\ \citenamefont
  {Strachan}(2014)}]{Avdoshenko:2014}%
  \BibitemOpen
  \bibfield  {author} {\bibinfo {author} {\bibfnamefont {S.~M.}\ \bibnamefont
  {Avdoshenko}}\ and\ \bibinfo {author} {\bibfnamefont {A.}~\bibnamefont
  {Strachan}},\ }\bibfield  {title} {\bibinfo {title} {High-temperature
  emissivity of silica, zirconia and samaria from ab initio simulations: role
  of defects and disorder},\ }\href
  {https://doi.org/10.1088/0965-0393/22/7/075004} {\bibfield  {journal}
  {\bibinfo  {journal} {Modelling and Simulation in Materials Science and
  Engineering}\ }\textbf {\bibinfo {volume} {22}},\ \bibinfo {pages} {075004}
  (\bibinfo {year} {2014})}\BibitemShut {NoStop}%
\bibitem [{\citenamefont {Harl}(2008)}]{Harl08}%
  \BibitemOpen
  \bibfield  {author} {\bibinfo {author} {\bibfnamefont {J.}~\bibnamefont
  {Harl}},\ }\emph {\bibinfo {title} {The linear response function in density
  functional theory}},\ \href {https://doi.org/10.25365/thesis.2622} {\bibinfo
  {type} {{Ph.D.} thesis}},\ \bibinfo  {school} {University of Vienna, Austria}
  (\bibinfo {year} {2008})\BibitemShut {NoStop}%
\end{thebibliography}
%

\end{document}